\documentclass[longauth]{aa}  

\usepackage{graphicx}
\usepackage{txfonts}
\usepackage{url}
\usepackage{xcolor}
\usepackage{hyperref}
\hypersetup{
    colorlinks,
    linkcolor={blue},
    citecolor={blue},
    urlcolor={blue}
}

\begin{document}

\title{Resolving faint structures in the debris disk around TWA\,7\thanks{Based on observations made with ESO Telescopes at the Paranal Observatory under programs ID 095.C-0298, 097.C-0319, 098.C-0155, and 198.C-0209.}}
\subtitle{Tentative detections of an outer belt, a spiral arm, and a dusty cloud}

\author{
	J. Olofsson\inst{1,2,3} \and 
	R.~G. van Holstein\inst{4} \and 
	A. Boccaletti\inst{5} \and 
	M. Janson\inst{1,6} \and 
	P. Th\'ebault\inst{5} \and 
	R. Gratton\inst{7} \and 
	C. Lazzoni\inst{7,8} \and 
	Q. Kral\inst{5,9} \and 
	A. Bayo\inst{2,3} \and 
	H. Canovas\inst{10} \and 
	C. Caceres\inst{11,3} \and 
	C. Ginski\inst{4} \and 
	C. Pinte\inst{12,13} \and 
	R. Asensio-Torres\inst{6} \and 
	G. Chauvin\inst{12,14} \and 
	S. Desidera\inst{7} \and 
	Th. Henning\inst{1} \and 
	M. Langlois\inst{15,16} \and 
	J. Milli\inst{17} \and
	J.~E. Schlieder\inst{18,1} \and
	M.~R. Schreiber\inst{2,3} \and
	J.-C. Augereau\inst{12} \and
	M. Bonnefoy\inst{12} \and
	E. Buenzli\inst{19} \and
	W. Brandner\inst{1} \and
	S. Durkan\inst{20,6} \and
	N. Engler\inst{19} \and
	M. Feldt\inst{1} \and
	N. Godoy\inst{2,3} \and
	C. Grady\inst{21} \and
	J. Hagelberg\inst{12} \and
	A.-M. Lagrange\inst{12} \and
	J. Lannier\inst{12} \and
	R. Ligi\inst{22} \and
	A.-L. Maire\inst{1} \and
	D. Mawet\inst{23,24} \and
	F. M\'enard\inst{12} \and
	D. Mesa\inst{7,25} \and
	D. Mouillet\inst{12} \and
	S. Peretti\inst{26} \and
	C. Perrot\inst{5} \and
	G. Salter\inst{16} \and
	T. Schmidt\inst{5} \and
	E. Sissa\inst{7} \and
	C. Thalmann\inst{19} \and
	A. Vigan\inst{16} \and
	L. Abe\inst{27} \and
	P. Feautrier\inst{12} \and
	D. Le Mignant\inst{16} \and
	T. Moulin\inst{12} \and
	A. Pavlov\inst{1} \and
	P. Rabou\inst{12} \and
	G. Rousset\inst{5} \and
	A. Roux\inst{12}
}

\institute{
	Max Planck Institut f\"ur Astronomie, K\"onigstuhl 17, 69117 Heidelberg, Germany
	\and
	Instituto de F\'isica y Astronom\'ia, Facultad de Ciencias, Universidad de Valpara\'iso, Av. Gran Breta\~na 1111, Playa Ancha, Valpara\'iso, Chile\\\email{johan.olofsson@uv.cl}
	\and
	N\'ucleo Milenio Formaci\'on Planetaria - NPF, Universidad de Valpara\'iso, Av. Gran Breta\~na 1111, Valpara\'iso, Chile
	\and
	Leiden Observatory, Leiden University, P.O. Box 9513, 2300 RA Leiden, The Netherlands
	\and
	LESIA, Observatoire de Paris, Universit\'e PSL, CNRS, Sorbonne Universit\'e, Univ. Paris Diderot, Sorbonne Paris Cit\'e, 5 place Jules Janssen, 92195 Meudon, France
	\and
	Department of Astronomy, Stockholm University, 106 91 Stockholm, Sweden
	\and
	INAF - Osservatorio Astronomico di Padova, Vicolo dell'Osservatorio 5, I-35122 Padova, Italy
	\and
	Dipartimento di Fisica a Astronomia ``G. Galilei'', Universita' di Padova, Via Marzolo, 8, 35121 Padova, Italy
	\and
	Institute of Astronomy, University of Cambridge, Madingley Road, Cambridge CB3 0HA, UK
	\and 
	European Space Astronomy Centre (ESA), Camino Bajo del Castillo s/n, 28692, Villanueva de la Ca\~nada, Madrid, Spain
	\and
	Departamento de Ciencias Fisicas, Facultad de Ciencias Exactas, Universidad Andres Bello. Av. Fernandez Concha 700, Las Condes, Santiago, Chile
	\and
	Univ. Grenoble Alpes, CNRS, IPAG, F-38000 Grenoble, France
	\and
	Monash Centre for Astrophysics (MoCA) and School of Physics and Astronomy, Monash University, Clayton Vic 3800, Australia
	\and
	Unidad Mixta Internacional Franco-Chilena de Astronom\'{i}a, CNRS/INSU UMI 3386 and Departamento de Astronom\'{i}a, Universidad de Chile, Casilla 36-D, Santiago, Chile
	\and
	CRAL, UMR 5574, CNRS/ENS-L/Universit\'e Lyon 1, 9 av. Ch. Andr\'e, F-69561 Saint-Genis-Laval, France
	\and
	Aix Marseille Univ., CNRS, LAM, Laboratoire d'Astrophysique de Marseille, Marseille, France
	\and 
	European Southern Observatory, Alonso de C\'ordova 3107, Casilla 19001 Vitacura, Santiago 19, Chile
	\and
	Exoplanets and Stellar Astrophysics Laboratory, Code 667, NASA Goddard Space Flight Center, Greenbelt, MD 20770, USA
	\and
	Institute for Particle Physics and Astrophysics, ETH Zurich, Wolfgang-Pauli-Strasse 27, 8093 Zurich, Switzerland
	\and
	Astrophysics Research Centre, Queen's University Belfast, Belfast, Northern Ireland, UK
	\and
	Eureka Scientific, 2452 Delmer, Suite 100, Oakland CA 96002, USA
	\and
	INAF-Osservatorio Astronomico di Brera, Via E. Bianchi 46, I-23807 Merate, Italy.
	\and
	Department of Astrophysics, California Institute of Technology, 1200 E. California Boulevard, Pasadena, CA 91101, USA
	\and
	Jet Propulsion Laboratory, California Institute of Technology, Pasadena, CA 91109, USA
	\and
	INCT, Universidad De Atacama, calle Copayapu 485, Copiap\'{o}, Atacama, Chile
	\and 
	D\'epartement d'Astronomie, Universit\'e de Gen\`eve, 51 chemin des Maillettes, 1290, Versoix, Switzerland
	\and
	Universit\'e C\^ote d'Azur, OCA, CNRS, Lagrange, France
}

\date{}

\abstract
{Debris disks are the intrinsic by-products of the star and planet formation processes. Most likely due to instrumental limitations and their natural faintness, little is known about debris disks around low mass stars, especially when it comes to spatially resolved observations.}
{We present new VLT/SPHERE IRDIS Dual-Polarization Imaging (DPI) observations in which we detect the dust ring around the M2 spectral type star TWA\,7. Combined with additional Angular Differential Imaging observations we aim at a fine characterization of the debris disk and setting constraints on the presence of low-mass planets.}
{We model the SPHERE DPI observations and constrain the location of the small dust grains, as well as the spectral energy distribution of the debris disk, using the results inferred from the observations, and perform simple N-body simulations.}
{We find that the dust density distribution peaks at $\sim0.72\arcsec$ ($25$\,au), with a very shallow outer power-law slope, and that the disk has an inclination of $\sim 13^{\circ}$ with a position angle of $\sim 91^{\circ}$ East of North. We also report low signal-to-noise detections of an outer belt at a distance of $\sim 1.5\arcsec$ ($\sim 52$\,au) from the star, of a spiral arm in the Southern side of the star, and of a possible dusty clump at $0.11\arcsec$. These findings seem to persist over timescales of at least a year. Using the intensity images, we do not detect any planets in the close vicinity of the star, but the sensitivity reaches Jovian planet mass upper limits. We find that the SED is best reproduced with an inner disk at $\sim0.2\arcsec$ ($\sim 7$\,au) and another belt at $0.72\arcsec$ ($25$\,au).}
{We report the detections of several unexpected features in the disk around TWA\,7. A yet undetected $100$\,M$_\oplus$ planet with a semi-major axis at $20-30$\,au could possibly explain the outer belt as well as the spiral arm. We conclude that stellar winds are unlikely to be responsible for the spiral arm.}


\maketitle

\section{Introduction}

Debris disks are the leftovers of the star and planet formation processes. Once the original gaseous disk has been dissipated, on a timescale of a few Myr (\citealp{Hernandez2007}), only planetesimals (and possibly already formed planets) remain. These large, unseen bodies will continuously release a population of small $\muup$m-sized dust grains that can be observed in scattered light images (see reviews by \citealp{Wyatt2008,Krivov2010,Matthews2014}). Debris disks are detected around about $20$\,\% of F, G, and K type stars (\citealp{Eiroa2013,MontesinosB2016}), and are more often detected around early type stars ($25-33$\,\% for A-type stars, \citealp{Su2006,Thureau2014}) rather than late spectral type stars (\citealp{Plavchan2009}). As discussed in \citet{Morey2014}, this possible trend may be an observational bias, as facilities such as \textit{Spitzer} and \textit{Herschel} were not sensitive enough to efficiently detect the cold dust around low-mass stars. Nevertheless, knowing that they harbor protoplanetary disks in their youth (e.g., \citealp{Pascucci2016}), which remains true even for the lowest mass object, free-floating planets (\citealp{Bayo2017}), and that those disks are capable of forming planets (e.g., \citealp{Chauvin2004,Gillon2017,Dressing2015}), it is likely that low-mass stars also harbor debris disks. But their in-depth characterization is greatly hindered by low-number statistics, especially when it comes to spatially resolved observations. The handful of exceptions being AU\,Mic (\citealp{Liu2004}), TWA\,7, TWA\,25 (\citealp{Choquet2016}), and GJ\,581 (\citealp{Lestrade2012}).

We recently observed TWA\,7 with the SPHERE instrument (\citealp{Beuzit2008}) at the \textit{Very Large Telescope}, and discovered unexpected faint structures in the circumstellar disk. TWA\,7 is an M2 star, at a distance of $34.5 \pm 2.5$\,pc (\citealp{Ducourant2014}, TWA\,7 is not in the Tycho-Gaia Astrometric Solution catalog) and is a member of the TW Hydra moving group (\citealp{Webb1999}). Membership to this young association would imply an age of $10\pm3$\,Myr (\citealp{Bell2015}, older than the $4$\,Myr proposed by \citealp{Herczeg2014}). Infrared (IR hereafter) excess was reported by \citet{Low2005}, using \textit{Spitzer}/MIPS data at $24$ and $70$\,$\muup$m, and since then, the spectral energy distribution (SED) has been studied by several authors (e.g., \citealp{Low2005,Matthews2007,Riviere2013}). Using SCUBA observations, \citet{Matthews2007} concluded that models with a single dust temperature were not capable of reproducing the SED, and suggested either a distribution of grain sizes, or an extended disk (overall, contributions from dust grains at different temperatures). \citet{Riviere2013}, using \textit{Herschel}/PACS observations reached similar conclusions; a single temperature modified blackbody cannot successfully match the entire SED of TWA\,7. They postulated that instead of an extended disk or a grain size distribution, the disk consists of two spatially separated dust belts, one at $\sim 38$\,au and another at $\sim 75$\,au. They did not detect the $[\ion{O}{i}]$ emission in the PACS spectroscopic observations, and regarding CO, \citealp{Doppmann2017} did not detect any emission lines in high-resolution $4.7$\,$\muup$m spectroscopic NIRSPEC observations. More recently, \citet{Holland2017} also modeled the SED of TWA\,7, with additional unresolved SCUBA-2 observations, and the best fit model suggests that there are two dust rings around the central star, one at $2.5$\,au and the other one at $\sim 49$\,au. The differences underline the unfortunate degeneracies when modeling unresolved photometric observations (the SCUBA-2 image yields an upper limit for the radius of $380$\,au), and the crucial need for spatially resolved observations. \citet{Choquet2016} presented the first resolved observations of the disk around TWA\,7, using the \textit{Hubble Space Telescope}/NICMOS instrument. They performed forward modeling of the disk, but could not fully constrain the dust distribution in the innermost regions of the disk. The authors present two equivalent solutions; a dust ring peaking at $35\pm3$\,au or a continuous disk that extends closer to the star and starts decreasing at $45\pm5$\,au. No additional belt was detected in these observations, most likely due to an overall low S/N. 

In this paper, we present a set of high angular resolution observations with SPHERE and NACO at the VLT (\citealp{Lenzen2003,Rousset2003}) to better characterize the disk around TWA\,7, one of the four resolved debris disks around an M type star. We first present the observations and data reduction. In section\,\ref{sec:modeling} we model the Dual-Polarization Imaging (DPI) observations, and the SED in section\,\ref{sec:sed}. We discuss our results in section\,\ref{sec:discussion} before concluding.

\section{Observations, data reduction, and results}

Table\,\ref{tab:log} summarizes the observing conditions and different instrumental setups for the observations used in this paper.

\begin{table*}
\centering
\caption{Log for the VLT/SPHERE and NACO observations. The integration time corresponds to the on-source time.}
\label{tab:log}
\begin{tabular}{@{}lclcccc@{}}
\hline\hline
Observing date   & Prog. ID & Instrument Mode & Filter & Seeing & Coherence time & Integration\\
$[$YYYY-MM-DD$]$ &          &                 &        & $[\arcsec]$ & $[$ms$]$ & $[$hrs$]$ \\
\hline
2015-05-09 & 095.C-0298 & SPHERE/IRDIFS      & $H2H3$/$YJ$   & 1.31 & 2.0 & 1.07 / 1.07  \\
2016-04-28 & 097.C-0319 & SPHERE/IRDIS DPI   & $B\_J$        & 0.97 & 2.7 & 0.36  \\
2017-01-14 & 098.C-0155 & NACO (AGPM)        & $L'$          & 0.75 & 9.5 & 0.40  \\
2017-02-07 & 198.C-0209 & SPHERE/IRDIFS      & $H2H3$/$YJ$   & 0.54 & 5.6 & 1.45 / 1.37  \\
2017-03-20 & 198.C-0209 & SPHERE/IRDIS DPI   & $B\_H$        & 0.85 & 5.3 & 0.78  \\
\hline
\end{tabular}
\end{table*}

\subsection{SPHERE/IRDIS DPI observations}\label{sec:data_DPI}

\begin{figure*}
\centering
\includegraphics[width=\hsize]{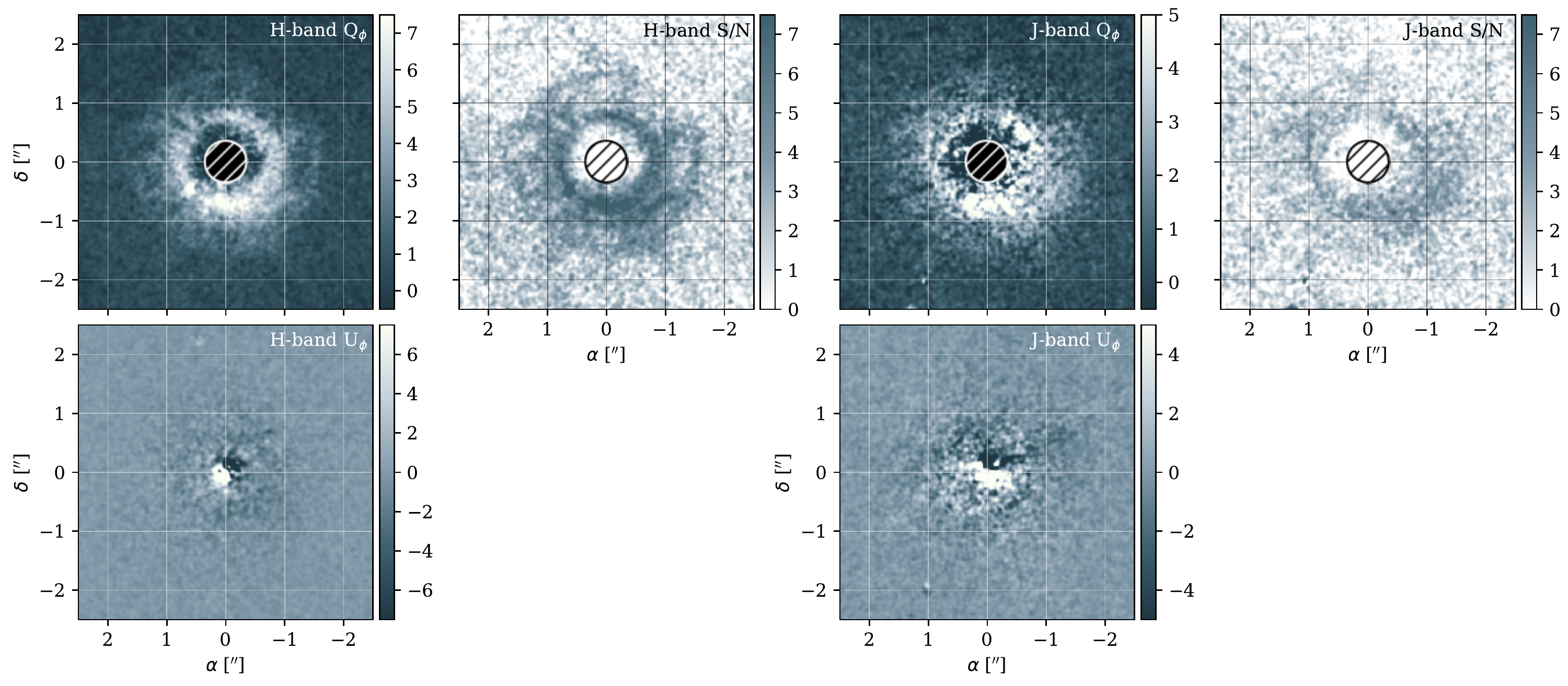}
\caption{\textit{Top panels:} $Q_\phi$ images and signal to noise maps for observations performed in P98 (leftmost panels, in $H$-band) and in P97 (rightmost panels, in $J$-band). \textit{Bottom panels:} $U_\phi$ images. North is up, East is left. Images have been convolved with a 2D gaussian of 2 pixels width, and the color scale is linear (in arbitrary units). Except for the $U_\phi$ images, the inner $0.35\arcsec$ in radius are masked (larger than the size of the coronagraph).}
\label{fig:dpi_data}
\end{figure*}

TWA\,7 was observed twice with SPHERE in DPI mode (\citealp{Langlois2014}) with the IRDIS instrument (\citealp{Dohlen2008}, with a pixel size of $12.26$\,mas). Observations were taken in P97 (open time program, 097.C-0319(A)) in $J$-band, and in P98 (guaranteed time observations, 198.C-0209(F)), in $H$-band, with a Detector Integration Time of $64$\,sec in both cases, and we used the N\_ALC\_YJH\_S coronagraph ($185$\,milli-arcsec in diameter). The observations are obtained in field-tracking mode and the light is split into two perpendicular polarization directions. By rotating the half-wave plate (HWP) at different switch angles ($0^{\circ}$, $22.5^{\circ}$, $45^{\circ}$, and $67.5^{\circ}$) one can construct the Stokes $Q$ and $U$ images. Our observations contain $5$ and $11$ full HWP cycles in period 97 and 98, respectively. During the P98 observations, as the observations were performed in $H$-band, we also included a derotator angle offset of $-220^{\circ}$ to avoid loss of polarization signal due to an expected significant drop in polarimetric efficiency (de Boer et al., in prep.; van Holstein et al., in prep.).

Basic data reduction was performed using the SPHERE Data Reduction Handling pipeline (\citealp{Pavlov2008}), for the background subtraction, flat field correction and centering of the frames. To estimate the location of the star behind the coronagraph, we used the two centering frames taken before and after the DPI sequence. After finding the location of the star we averaged the two positions along the two axis and used those values to re-center all the frames. For the P97 dataset, the difference between the two positions only differs by less than $0.3$\,pixels in both directions. For P98, the positions differ by $0.3$\,pixel along the x-axis and $0.9$\,pixel along the y direction. Afterwards, the instrumental polarization and instrument-induced cross-talk in the optical path of the telescope and the instrument were corrected for with the detailed instrument model and pipeline presented in van Holstein et al. (in prep) and \citet[][see also \citealp{Pohl2017} and \citealp{Canovas2018}]{Holstein2017}. Subsequently, we computed the azimuthal Stokes images ($Q_{\phi}$ and $U_{\phi}$, see \citealp{Schmid2006,Avenhaus2014}) by combining the frames altogether in their raw orientation before rotating them to account for the derotator angle offset and align the North up. Doing so avoid losing signal due to several interpolation when de-rotating individual frames.

To try to increase the signal from the disk, given its overall faintness, we also attempted to perform frame selection. The motivation was to discard frames for which the central star slightly moved under the coronagraph. In the end, this did not provide clear improvements, as it is more important to have as much flux as possible rather than remove a handful of inhomogeneous frames.

Figure\,\ref{fig:dpi_data} shows the final reduced images for both periods. On the top panels, from left to right are shown: the $Q_{\phi}$ image and signal-to-noise (S/N) map for $H$-band and $J$-band, respectively. On the bottom panels, the $U_\phi$ images are also shown. All images have been convolved with a 2D Gaussian with a $2$\,pixel standard deviation, to reduce the shot noise. The uncertainties used to generate the S/N maps were derived from the $U_{\phi}$ image by measuring the standard deviation in concentric, $2$\,pixel wide, annuli. Consequently, the estimated uncertainties are correlated by the convolution, and furthermore, the azimuthal information is lost. We only have a radial-dependent estimate of the uncertainties. Assuming that there are no multiple scattering events in the disk (a reasonable assumption for optically thin debris disks and the fact that the disk is seen at low inclination), the $U_{\phi}$ image should not contain any astrophysical signal (\citealp{Canovas2015}), which is verified in our observations. With observations performed in field-tracking mode, one could attempt to perform Reference star Differential Imaging to retrieve signal from the disk in total intensity. However, given how faint the signal is, we did not attempt such an approach that requires finding the best suited point spread functions to match the observed ones.

\begin{figure}
\centering
\includegraphics[width=\hsize]{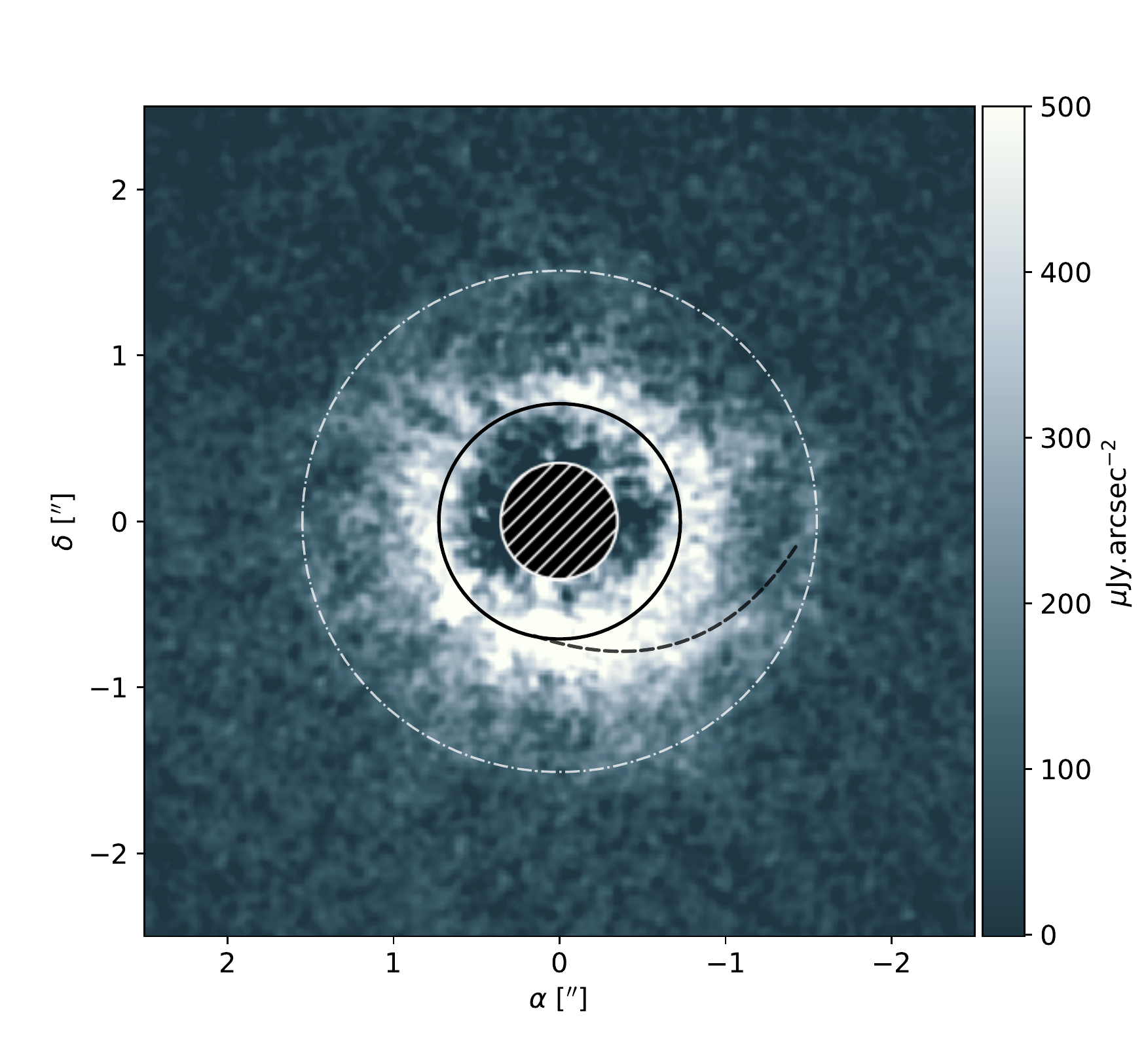}
\caption{Flux calibrated SPHERE/IRDIS DPI $H$-band observations ($Q_{\phi}$ image, in $\muup$Jy.arcsec$^{2}$), where we highlight the locations of the main and secondary belts, as well as the tentative spiral pattern.}
\label{fig:outline}
\end{figure}

Several things are to be noted from inspecting the images. First, the disk is clearly detected in both epochs, at the $\sim 5-7\sigma$ and $4\sigma$ levels in the $H$ and $J$-band datasets respectively. The fainter signal obtained in $J$-band can be explained by the shorter total integration time. Second, there seems to be a faint outer ring (at about $1.5\arcsec$), which appears stronger in the $H$-band dataset ($3-4\sigma$ level) but can be seen in the North-East direction in the $J$-band observations. Finally, there is a tentative detection of a spiral arm in the South-West direction. This spiral arm seems to be detected in both datasets, at the $4-5\sigma$ level, and cannot be attributed to instrumental polarization effects: it is seen both in $J$- and $H$-band, and with different derotator offsets for the detector. Figure\,\ref{fig:outline} highlights the features of interest we will further discuss in this paper.

The $H$-band observations shown in Figure\,\ref{fig:outline} are also photometrically calibrated. The calibration was done using the off-axis frame taken during the observations, with a neutral density filter (ND\_2.0). We determined the flux in the off-axis frame by performing aperture photometry, with a circular aperture ($50$\,pixels in radius), after having subtracted the median background contribution (estimated within an annulus between $50$ and $60$\,pixels in radius). The stellar flux was then corrected for the differences in Detector Integration Time, and the neutral density filter was accounted for. The Johnson $H$-band magnitude of TWA\,7 is $7.125$ corresponding to a flux of $1.46$\,Jy at $1.65$\,$\muup$m. For each pixel of the $Q_{\phi}$ image of Fig.\,\ref{fig:outline}, we multiplied the number of counts by the stellar flux in units of $\muup$Jy, divided by the stellar counts measured from the off-axis frame, and divided by the pixel scale ($0.01226\arcsec$) squared.

Nonetheless, for the rest of the analysis, we will work on the native $Q_{\phi}$ image (without the photometric calibration). It is extremely challenging to properly estimate the uncertainties of the photometrically calibrated image due to possible PSF variations during the observing sequence (for instance because of seeing variations). Since proper uncertainties are required to perform the modeling, we therefore chose to use the native $Q_{\phi}$ and $U_{\phi}$ images.

\subsection{SPHERE/IRDIFS ADI observations}

TWA\,7 is part of SHINE, the \textit{``SpHere INfrared survey for Exoplanets''} (\citealp{Chauvin2017}) and was observed twice with SPHERE in 2015-05-09 (095.C-0298(A)) and 2017-02-07 (198.C-209(D)). As the second sequence was of better quality, here we will mostly focus on the data from 2017 (but will use the 2015 dataset in section\,\ref{sec:coincidence}). Observations were carried out in pupil tracking (to allow for Angular Differential Imaging, ADI, \citealp{Marois2006}) with the IRDIFS mode and we also used the N\_ALC\_YJH\_S coronagraph ($185$\,milli-arcsec in diameter). The IRDIFS mode combines IRDIS the NIR dual-band camera using the $H2H3$ filter ($1.593$, $1.667$\,$\muup$m, in Dual Band Imaging, \citealp{Vigan2010}) and IFS (\citealp{Claudi2008}, with a pixel size of $7.46$\,mas) the NIR integral field spectrograph in $YJ$ ($0.95-1.35$\,$\muup$m, $R\sim54$)
 
The observing sequence consists of a PSF short observation to calibrate the photometry, a first ``starcenter'' frame with waffle imprinted on the deformable mirror shape, and a series of deep coronagraphic images ($163$ frames of $32$\,sec, a field rotation $\Delta\theta = 90.8^{\circ}$). The PSF and starcenter measurements are repeated at the end of the sequence together with sky observations. The location of the star under the coronagraph is determined through the diffracted satellite spots of the deformable mirror (e.g., \citealp{Langlois2013}). By fitting a gaussian function to each of the four spot, the intersecting point is determined by joining two opposite spots.

Similarly to DPI, the IRDIFS data were processed with the SPHERE Data Reduction Handling pipeline. Distortion and True North corrections are provided in \citet{Maire2016}. Star light suppression using ADI or in combination with angular spectral differential imaging (ASDI, \citealp{Mesa2015}) was achieved with a dedicated IDL-based pipeline, SpeCal (Galicher et al., submitted) implemented at the SPHERE Data Center\footnote{\url{http://sphere.osug.fr}} (\citealp{Delorme2018}). The debris disk was not detected in these observations, not surprisingly given its low inclination which leads to significant self-subtraction of the astrophysical signal (\citealp{Milli2012}). The constraints on low-mass planets (based on PCA with $10$ components, \citealp{Soummer2012}) are further discussed in section\,\ref{sec:planets}. The reduced images for IRDIS and IFS are shown in Fig.\,\ref{fig:irdis_image} and \ref{fig:ifs_image}, respectively.

\subsection{NACO ADI observations}

TWA\,7 was observed with VLT/NACO in the $L^{\prime}$-band on 2017-01-14 (program 098.C-0155(A)). The observations, with a pixel size of $27$\,mas, made use of the Annular Groove Phase Mask (AGPM, with an inner working angle of $\sim0.09\arcsec$ \citealp{Mawet2005,Mawet2013}) coronagraph in order to optimize the contrast at small separations from the star, and additionally were performed in pupil stabilized mode to facilitate the implementation of ADI for further contrast improvement. Sky frames were interspersed at regular intervals among the coronagraphic frames. During the observations, the star is centered behind the coronagraph manually and the centering is evaluated by the observer, but appeared very stable during the sequence. In total $92$ coronagraphic frames and $11$ sky frames were obtained, but since conditions were poor during the beginning of the observations, we excluded a substantial amount of frames and ended up with $57$ high quality coronagraphic frames. Each frame consisted of $126$ exposures with $0.2$\,s integration time each, so the effective useful on-target integration time was $24$ minutes. The total field rotation from first to last usable frame was $51^{\circ}$.

We reduced the NACO data with a custom pipeline in IDL, building on a previous work in this wavelength range \citep{Janson2008}. A flat field frame was acquired from exposures of the thermal background at different integration times, and the sky was estimated individually for each coronagraphic frame by interpolating in time between the sky frames. Apart from accommodating changes in the thermal background, this procedure also aids in removing multiple ghost features across the field in the raw frames, probably caused by internal reflections from the AGPM mask, which displayed a small degree of motion from the start to the end of the observations. From visual inspection of the residual PSF pattern, it was clear that the star had been very stably placed behind the mask, with $<$1 pixel drifts across the sequences.

PSF subtraction was performed using a LOCI procedure \citep{Lafreniere2007}. Since the residuals in the coronagraphic images are faint relative to the background, we only perform the optimization in an annulus between $0.4\arcsec$ and $0.8\arcsec$ separation. The reduction is quite conservative and maintains a $>$90\% throughput at all separations. An unsaturated image of TWA\,7 itself was meant to be taken for flux calibration purposes, but due to an execution error, no usable files were available. Instead, we calibrate the flux based on the thermal background level as compared to the corresponding flux level provided by the NACO exposure time calculator, which is quite stable against ambient conditions. We did not detect the disk in this dataset either, and the reduced image is presented in Fig.\,\ref{fig:naco_image}. The constraints brought by this dataset are presented in Section\,\ref{sec:planets}.

\section{Modeling of the DPI data}
\label{sec:modeling}

Because of the better S/N in the $H$-band observations, we perform the modeling on this dataset first and will then confront our best fit model to the $J$-band dataset to check for possible inconsistencies.

\subsection{Modeling strategy}

The modeling is performed using the same code as the one described in \citet{Olofsson2016}, which can produce synthetic images in scattered and polarized light. The only difference is that we implemented a parametric polarized phase function instead of using the Mie theory to compute the $S_{\mathrm{12}}$ element of the M\"uller matrix. In \citet{Olofsson2016} we fixed the minimum and maximum grain sizes ($s_{\mathrm{min}}$ and $s_{\mathrm{max}}$, respectively) when modeling the DPI data of HD\,61005, using the polarized phase function as a prior (which was calculated using the Mie theory for given $s_{\mathrm{min}}$, $s_{\mathrm{max}}$, and dust composition). In the case of TWA\,7 given the low inclination of the disk, we do not sample a lot of scattering angles, and therefore chose not to fix the polarized phase function. Nonetheless, computing the absorption/scattering efficiencies for each model is a serious bottleneck, and one possible solution is to interpolate those quantities for each grain size from a ``master'' opacity table. Another alternative solution, that we adopted for this paper as it gives a finer control on the shape of the phase function, is the one presented in \citet{Engler2017}. We assumed that the scattering phase function (the $S_{11}$ element of the M\"uller matrix) can be described by the analytical Henyey-Greenstein approximation, as
\begin{equation}
S_{11, \mathrm{HG}} = \frac{(1 - g^2)}{4\pi [1 + g^2 - 2g\mathrm{cos}(\theta)]^{3/2}},
\end{equation}
where $g$ is the anisotropic scattering factor ($-1 \leq g \leq 1$) and $\theta$ the scattering angle. We then approximate the polarized phase function as 
\begin{equation}
S_{12, \mathrm{HG}} = S_{11, \mathrm{HG}} \times \frac{1 - \mathrm{cos}^2(\theta)}{1 + \mathrm{cos}^2(\theta)}.
\end{equation}
For an isotropic scattering phase function ($g = 0$), the polarized phase function will peak at scattering angles of $90^{\circ}$ and for forward scattering cases $g \geq 0$ (backward scattering cases, $g \leq 0$) the polarized phase function will peak at angles short-wards of $90^{\circ}$ (long-wards of $90^{\circ}$, respectively). With this approximation, we assume that we are in the Rayleigh domain, a reasonable hypothesis given that small dust grains (compared to the wavelength) are likely to remain on bound orbits because of weak radiation pressure. Overall, this gives us finer control over the shape of the polarized phase function.

To estimate the goodness of fit of a given model, we compute the $\chi^2$ as the sum of the squared difference of the final image and the model, divided by the square of the uncertainties. To speed up the modeling process, we simply cropped the original images to a size of $300 \times 300$\,pixels ($3.68\arcsec \times 3.68\arcsec$). Because of the shot noise in the original data, we used the convolved $Q_{\phi}$ and $U_{\phi}$ images, and the uncertainties were estimated in concentric annuli as mentioned earlier. To model the IRDIS/DPI dataset, we considered the following free parameters: the inclination $i$, position angle $\phi$, the Henyey-Greenstein coefficient $g$, and the volumetric density distribution $n(r, z)$. The latter is defined by the inner and outer slopes of the density distribution $\alpha_{\mathrm{in}}$ ($> 0$) and $\alpha_{\mathrm{out}}$ ($< 0$), respectively, the reference radius $r_0$, and the standard deviation $h = r \times \mathrm{tan}(\psi)$ for the vertical distribution:
\begin{equation}\label{eqn:nr}
n(r, z) \propto \left[\left(\frac{r}{r_0}\right)^{-2\alpha_{\mathrm{in}}} + \left(\frac{r}{r_0}\right)^{-2\alpha_{\mathrm{out}}}\right]^{-1/2} \times \mathrm{e}^{-z^2/2h^2},
\end{equation}
where $\psi$ is the opening angle (we fixed $\psi = 0.05$, its exact value does not really matter for a face-on disk). Therefore, the surface density distribution $\Sigma (r) = \int_{-\infty}^{+\infty} n(r,z)\,\mathrm{d}z$ far beyond $r_0$ follows a slope of $\alpha_{\mathrm{out}} + 1$.

For each model, we first convolve the synthetic image with the same gaussian kernel as the DPI image (with a $2$\,pixel standard deviation), and we then scale the entire image by a factor $f_{\mathrm{flux}}$ that is estimated using a least squares method, to minimize the residuals, as 
\begin{equation}
    f_{\mathrm{flux}} = \cfrac{\sum \left( \cfrac{F_{\mathrm{obs}} \times F_{\mathrm{model}}}{\sigma^2} \right)}{\sum \left(\cfrac{F_{\mathrm{model}}}{\sigma}\right)^2},
\end{equation}
where $F_{\mathrm{obs}}$ is the observed image, $F_{\mathrm{model}}$ is the model image, and $\sigma$ the uncertainties. When computing the $\chi^2$, we only consider regions of the $Q_{\phi}$ image in which the disk is reliably detected; an annulus between $0.35\arcsec$ and $1.8\arcsec$. The $f_{\mathrm{flux}}$ parameter is not considered as an input free parameter in the modeling strategy as it is independently evaluated for each individual model. Because we are only modeling the main ring, the presence of the fainter outer ring may bias the determination of the $f_{\mathrm{flux}}$ scaling factor. Nonetheless, given its overall faintness we did not attempt to include a second ring in the modeling process. Finally, given that the derotation process slightly degraded the quality of the final $Q_{\phi}$ image, we performed the modeling on the original $Q_{\phi}$ and $U_{\phi}$ images (which do not include the derotation), and only derotated the final $Q_{\phi}$ image for display purposes.

The other parameters required for the modeling are related to the central star and the dust properties. These parameters are fixed. Concerning the stellar parameters, we assumed an effective temperature of $3500$\,K (\citealp{Matthews2007}) and a distance of $34.5$\,pc (\citealp{Ducourant2014}). For the dust properties we use the optical constant for astronomical silicates from \citet{Draine2003} and compute the absorption and scattering efficiencies using the Mie theory, for grains with sizes between $0.01$\,$\muup$m and $1$\,mm ($100$ different grain sizes). The grain size distribution is the canonical differential power-law d$n(s) \propto s^{-3.5}$d$s$, where $s$ is the grain size (\citealp{Dohnanyi1969}). Overall, the results do not really depend on the chosen dust properties as we are modeling a monochromatic image and that the true phase function is replaced by the Henyey-Greenstein approximation.

\subsection{Results}

\begin{table}
\centering
\caption{Best fit results for the modeling of the SPHERE observations.\label{tab:sphere}}
\begin{tabular}{@{}lccc@{}}
\hline\hline
Parameter               & Uniform prior  & $\sigma_{\mathrm{kde}}$ & Best-fit value \\
\hline
$r_0$ [au]              & $[15, 40]$     &  $0.1$                  & $25.0_{-1.1}^{+1.3}$ \\
$i$ [$^{\circ}$]        & $[1, 40]$      &  $0.05$                 & $13.1_{-2.6}^{+3.1}$ \\
$\phi$ [$^{\circ}$]     & $[85, 105]$    &  $0.1$                  & $91.0_{-8.9}^{+9.3}$ \\
$g$                     & $[0, 0.99]$    &  $0.01$                 & $0.63_{-0.21}^{+0.21}$ \\
$\alpha_{\mathrm{in}}$  & $[0.5, 10]$    &  $0.01$                 & $5.0_{-1.2}^{+1.5}$ \\
$\alpha_{\mathrm{out}}$ & $[-7.5, -0.5]$ &  $0.01$                 & $-1.5_{-0.2}^{+0.2}$ \\
\hline
\end{tabular}
\end{table}

\begin{figure*}
\centering
\includegraphics[width=\hsize]{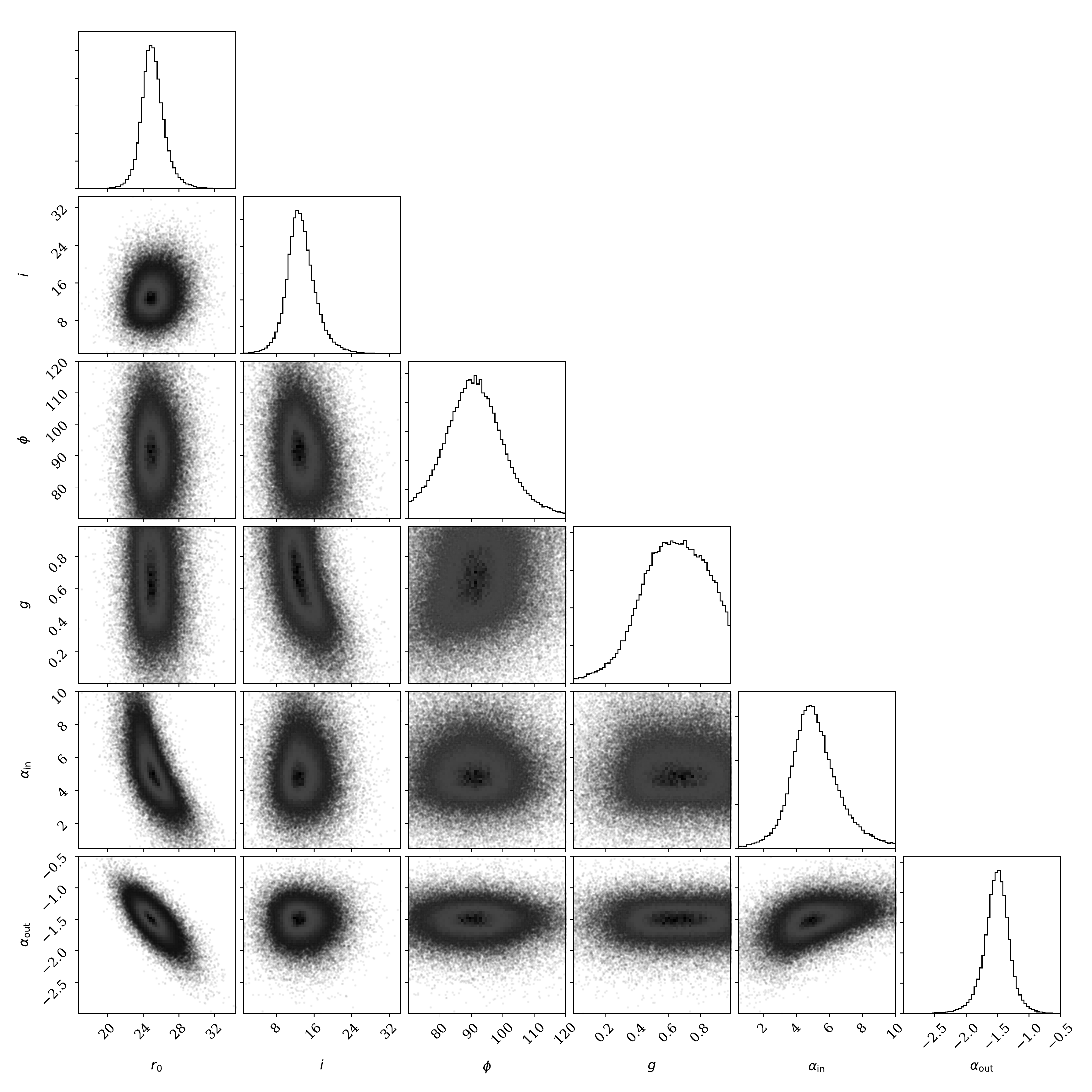}
\caption{Projected probability density distributions for the different parameters in the modeling as well as density plots.}
\label{fig:corner}
\end{figure*}

\begin{figure*}
\centering
\includegraphics[width=\hsize]{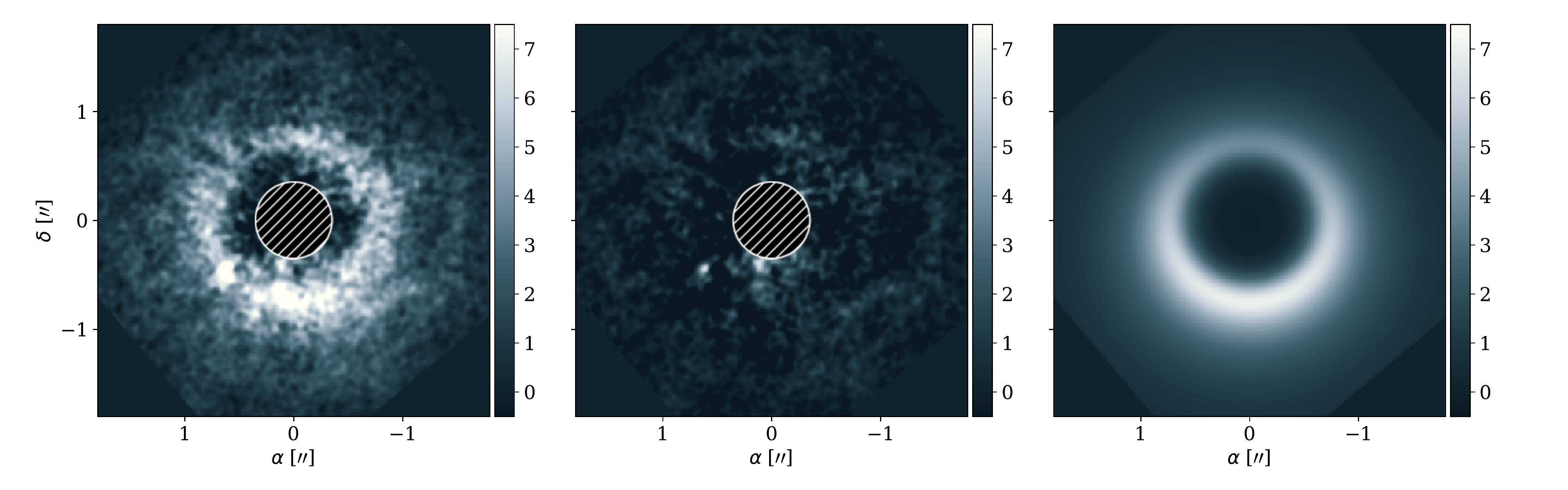}
\caption{{\it Left to right}: IRDIS/DPI data, residuals, best-fit model. All images have the same linear scaling. The modeling was done on the P98 dataset, which includes a derotator offset of $-220^{\circ}$. The images have been derotated afterwards for display purposes only. The point source that appears in the residual image is most likely an artifact of the adaptive optic system.}
\label{fig:results}
\end{figure*}

To find the best fitting solution, we used an affine invariant ensemble sampler (\texttt{emcee} package, \citealp{Foreman-Mackey2012}). We used 100 walkers, a burning phase of 600 steps and then iterated over 1\,500 steps for each of the walkers. At the end of the modeling, the mean acceptance fraction was $\sim 0.47$, and the maximum length for the auto-correlation time is $83$. Figure\,\ref{fig:corner} shows the one- and two-dimensional projections of the posterior probability distributions for all the free parameters in the modeling (using the \texttt{corner} package, \citealp{corner}). While all parameters appear to be well constrained, one can note that the Henyey-Greenstein coefficient $g$ is slightly degenerate with the inclination $i$. From the projected distributions, we estimated the most probable parameters for the best-fit model. We used a kernel density approach, with different kernel widths ($\sigma_{\mathrm{kde}}$), and estimated the most probable values as well as the $68$\,\% confidence intervals for each parameters. The values are reported in Table\,\ref{tab:sphere}. The most probable model is shown in Figure\,\ref{fig:results}, with from left to right, the DPI P98 dataset, the residuals, and the model (all panels have the same linear scaling). As a sanity check, we also compare our best fit model to the P97 dataset (in $J$-band) and found that the model can account for most of the detected signal, despite the lower S/N ratio (see Fig.\,\ref{fig:results_p97}).

\subsection{Inspecting the residuals}

\begin{figure*}
\centering
\includegraphics[width=0.45\hsize]{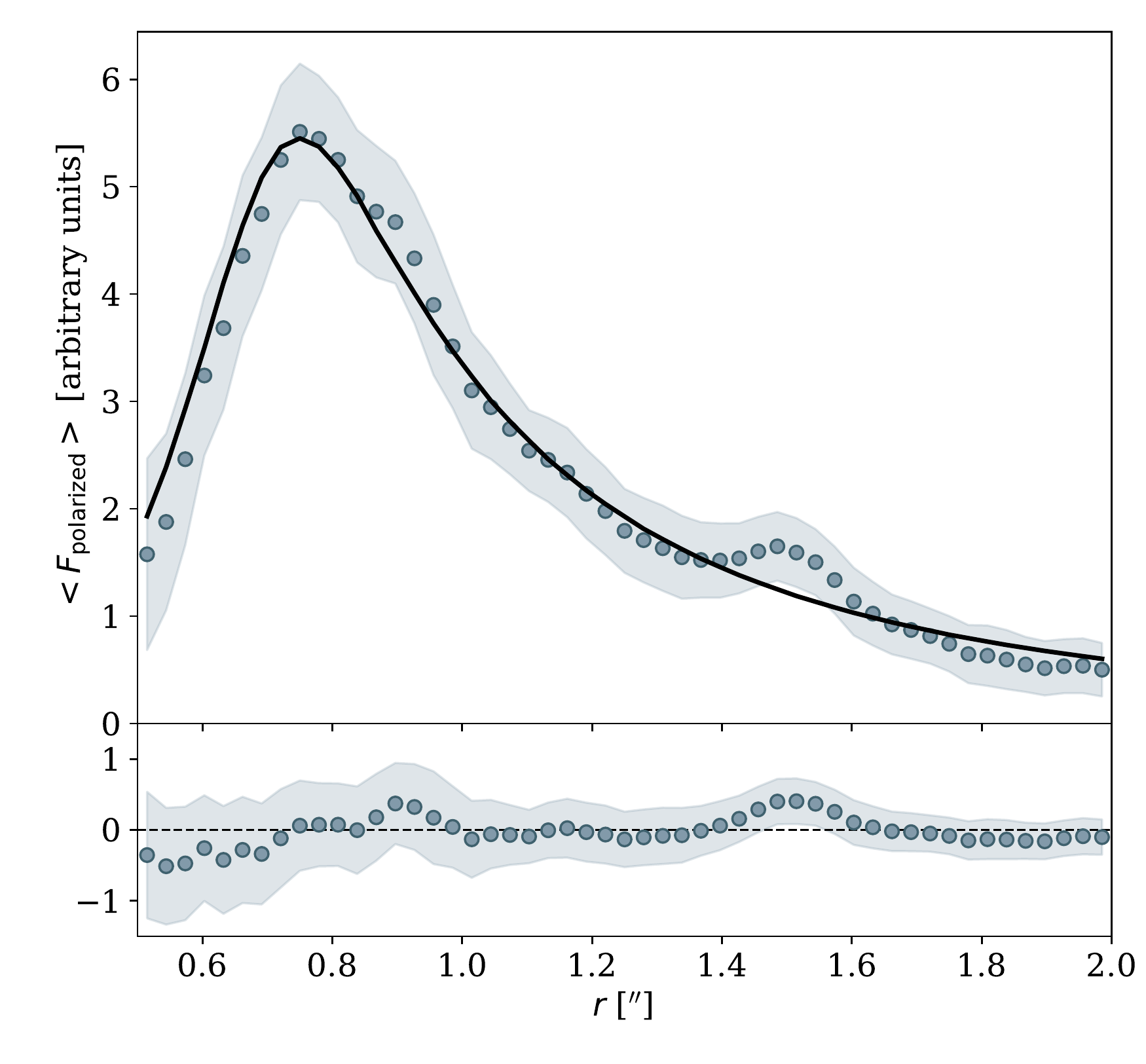}
\includegraphics[width=0.45\hsize]{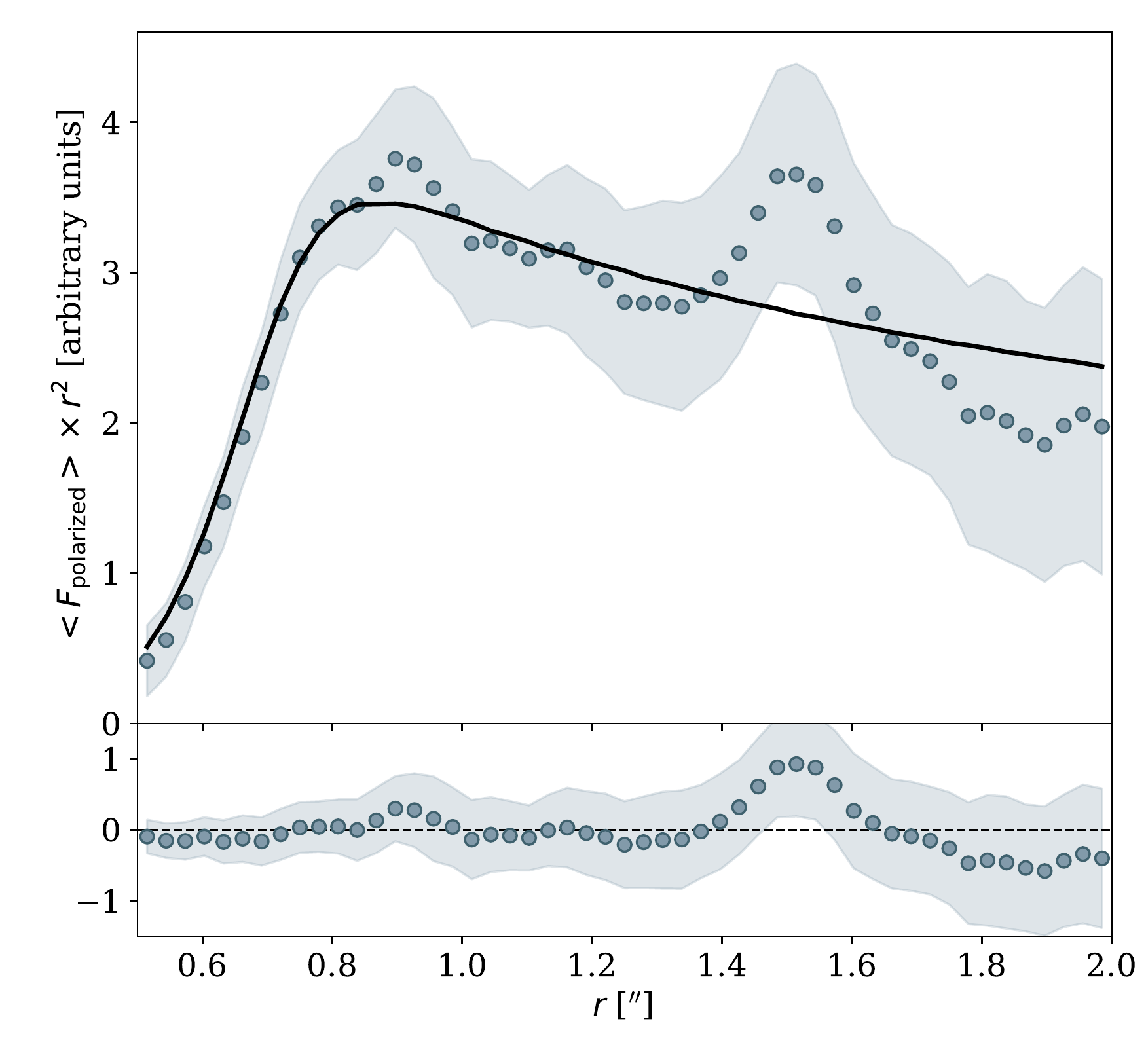}
\caption{Radial profiles extracted from the de-projected P98 observations (filled circles) and from the de-projected best fit model (black), for the polarized intensity and the polarized intensity times the distance squared (to compensate for illumination effects), left and right panels, respectively. The bottom panel shows the residuals once the best fit model is subtracted.}
\label{fig:radial}
\end{figure*}

In the middle panel of Fig.\,\ref{fig:results}, one can first remark the presence of a small ``speckle''-like feature on the South-East side. On the un-rotated images, this point falls on the right side of the star and is most likely an artifact of the adaptive optic system (a well-known speckle caused by a periodic pattern on the deformable mirror). In Section\,\ref{sec:planets} we discuss the detection limits of planets around TWA\,7.

To further investigate the presence of the second ring mentioned in Section\,\ref{sec:data_DPI}, we computed radial profiles, after de-projecting the $Q_{\phi}$ and $U_{\phi}$ images as well as the model. The profiles are calculated as the azimuthal mean in concentric annulii of $2$\,pixels width, while the uncertainties are the standard deviations divided by the square root of the number of pixels in the same annulus, in the U$_\phi$ image. The results and residuals are shown in Figure\,\ref{fig:radial}, for the polarized intensity and the intensity corrected for illumination effects (left and right, respectively). One can see that our model can reproduce the observations very well, except for the (tentative) secondary ring at $\sim1.5\arcsec$ (which is not included in the model). While the secondary ring appears faint in the observations, it becomes as bright as the primary ring when compensating for illumination effects as shown in the right panel of Figure\,\ref{fig:radial} where we multiplied the data by the square of the distance. As mentioned previously, this secondary ring is likely to introduce a bias in our modeling results as the model slightly over-predicts the flux at separations larger than $\sim1.6\arcsec$. Finally, the possible spiral arm is also seen in the residuals (middle panel of Fig.\,\ref{fig:results}), and can explain the ``shoulder'' in the radial profile at about $0.9\arcsec$. 

\section{Modeling the spectral energy distribution}
\label{sec:sed}

\begin{figure*}
\centering
\includegraphics[width=0.45\hsize]{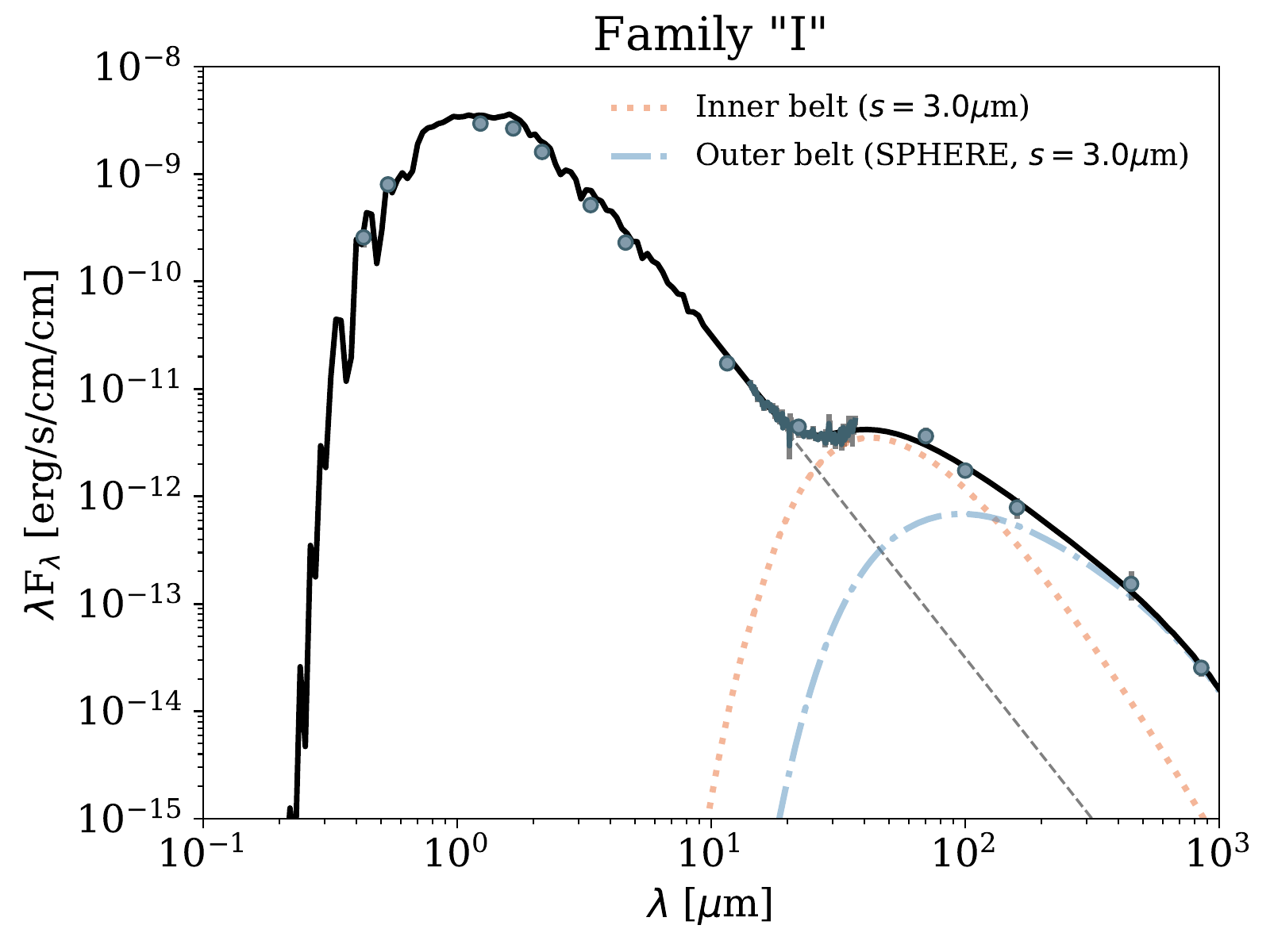}
\includegraphics[width=0.45\hsize]{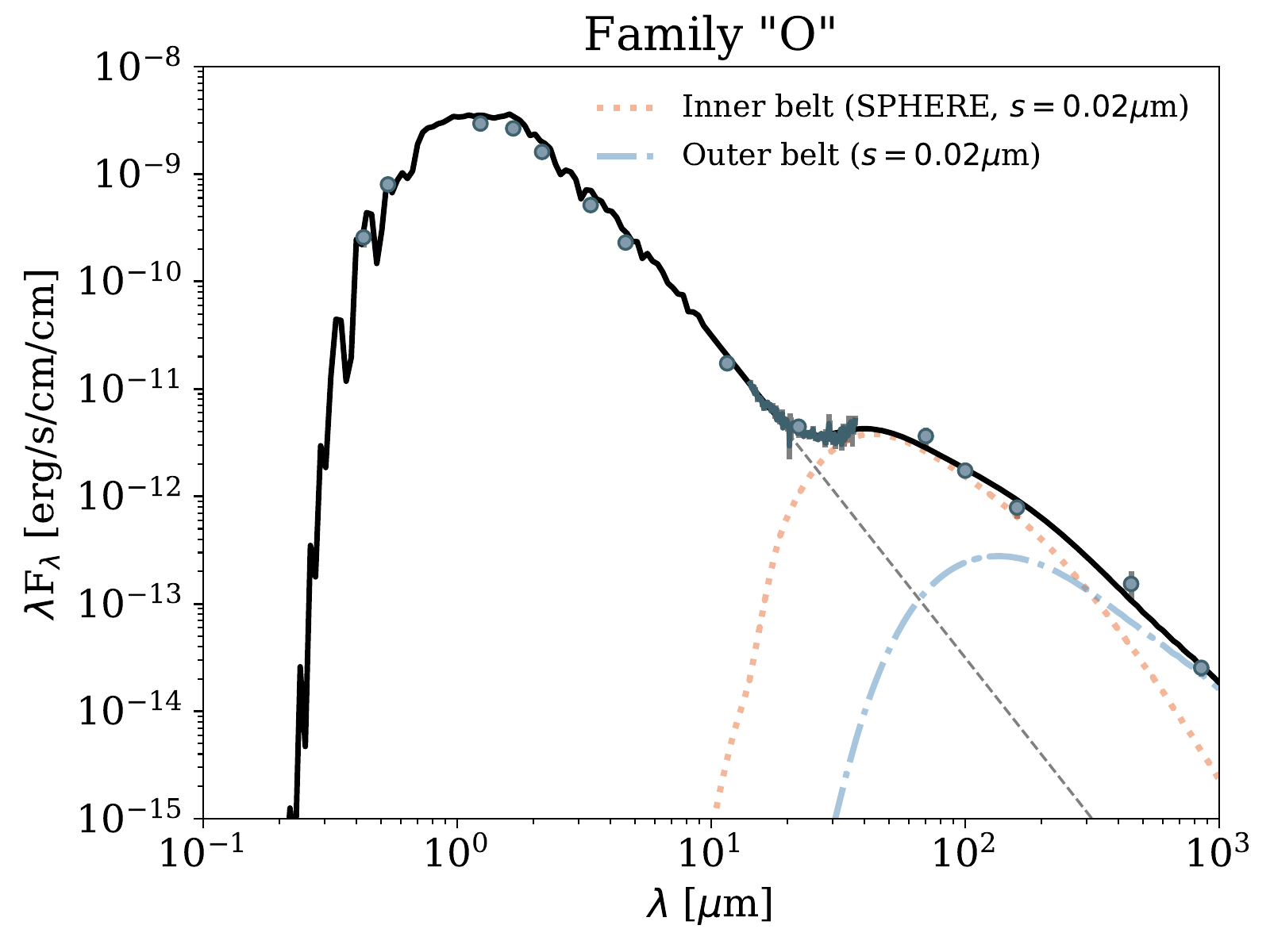}
\caption{Results of the SED modeling. \textit{Left:} we assume that the disk detected with SPHERE is the outer disk and that there is another dust belt close to the star. \textit{Right:} we assume that the disk detected with SPHERE is the inner belt, and that there is another belt farther out.}
\label{fig:sed}
\end{figure*}

We gathered optical and near-IR photometric points using the ``VO SED Analyzer'' tool (VOSA\footnote{\url{http://svo2.cab.inta-csic.es/theory/vosa/index.php}}, \citealp{Bayo2008}). We downloaded the low-resolution \textit{Spitzer}/IRS spectrum from the CASSIS\footnote{The Combined Atlas of Sources with Spitzer IRS Spectra (CASSIS) is a product of the IRS instrument team, supported by NASA and JPL.} database (\citealp{Lebouteiller2011}). The $70$, $100$, and $160$\,$\muup$m fluxes are taken from \citet{Riviere2013}, the $450$\,$\muup$m point from \citet{Matthews2007}, and the $850$\,$\muup$m measurement from \citet{Holland2017}. We assume the same stellar parameters as the ones mentioned earlier.

To model the SED, we proceed the exact same way as in \citet{Olofsson2016} and \citet{Feldt2017}, also making use of the \texttt{emcee} package. The volumetric density distribution follows the same expression as in Eqn.\,\ref{eqn:nr}, and we compute the thermal emission arising from the dust grains for wavelengths up to $1$\,mm before comparing the model to the observations. For the dust properties, we used the optical constant of astro-silicates (density $\rho = 3.5$\,g.cm$^{-3}$, \citealp{Draine2003}). We compute a master absorption efficiencies table with the Mie theory for sizes between $0.01$\,$\muup$m and $5$\,mm for $300$ sizes logarithmically spaced. In the modeling, $s_{\mathrm{min}}$ is a free parameter, and prior to computing the SED, we interpolate the absorption efficiencies, from the master table, at the adequate sizes, this time using $100$ sizes logarithmically spaced (with $s_{\mathrm{max}}$ still equal to $5$\,mm). As mentioned earlier, we cannot constrain the minimum grain size from the SPHERE/DPI images, as we parametrized the polarized phase function via the parameter $g$ (in section\,\ref{sec:ghg} we will discuss how well constrained this value actually is from the observations). To limit the number of free parameters, we fix the power-law of the grain size distribution to $p = -3.5$ and the inner slope of the density distribution $\alpha_{\mathrm{in}} = 5$ (close to what we infer from the modeling of the SPHERE data). To alleviate the known degeneracies of SED modeling (dependency between $s_{\mathrm{min}}$ and distance $r_0$), we use our modeling results and fix $r_0 = 25$\,au. 

\begin{table}
\centering
\caption{Best fit results for the modeling of the SED.}
\label{tab:sed}
\begin{tabular}{@{}lccc@{}}
\hline\hline
Parameter               & Uniform prior  & $\sigma_{\mathrm{kde}}$ & Best-fit value \\
\hline
\multicolumn{4}{c}{Additional inner belt - Family ``I''} \\
\hline
$s_{\mathrm{min}}$ [$\muup$m]           & $[0.1, 10]$    &  $0.01$                 & $3.0_{-0.2}^{+0.1}$ \\
$r_{0,\mathrm{inner}}$ [au]             & $[0.5, 15]$    &  $0.01$                 & $7.0_{-1.6}^{+0.4}$ \\
$\alpha_{\mathrm{out, inner}}$          & $[-20, -0.5]$  &  $0.01$                 & $-22.3_{-0.9}^{+5.5}$ \\
$\alpha_{\mathrm{out, outer}}$          & $[-20, -0.5]$  &  $0.01$                 & $-1.27_{-0.02}^{+0.02}$ \\
M$_{\mathrm{dust, inner}}$ [M$_\oplus$] &                &  $0.0001$               & $1.1_{-0.1}^{+0.4}\times 10^{-3}$ \\
M$_{\mathrm{dust, outer}}$ [M$_\oplus$] &                &  $0.01$                 & $0.57_{-0.03}^{+0.04}$ \\
\hline
\multicolumn{4}{c}{Additional outer belt - Family ``O''} \\
\hline
$s_{\mathrm{min}}$ [$\muup$m]           & $[0.01, 10]$   &  $0.01$                 & $0.02_{-0.01}^{+0.02}$ \\
$r_{0,\mathrm{outer}}$ [au]             & $[30, 300]$    &  $1.0$                  & $298_{-7}^{+2}$ \\
$\alpha_{\mathrm{out, inner}}$          & $[-20, -0.5]$  &  $0.01$                 & $-4.34_{-0.13}^{+0.09}$ \\
$\alpha_{\mathrm{out, outer}}$          & $[-20, -0.5]$  &  $0.01$                 & $-1.45_{-0.03}^{+0.02}$ \\
M$_{\mathrm{dust, inner}}$ [M$_\oplus$] &                &  $0.0001$               & $1.2_{-0.4}^{+0.4} \times 10^{-2}$ \\
M$_{\mathrm{dust, outer}}$ [M$_\oplus$] &                &  $0.1$                  & $4.4_{-0.3}^{+0.3}$ \\
\hline
\end{tabular}
\end{table}

We attempted to fit the entire SED using only one dust belt (free parameters being $s_{\mathrm{min}}$ and $\alpha_{\mathrm{out}}$, the dust mass being scaled to best match the observed fluxes), but could not reach a decent match to the photometric measurements (see Section\,5.1 of \citealp{Olofsson2016} for how the goodness of fit is estimated). The model could match the mid-IR turn-off point as well as the photometric point up to $160$\,$\muup$m, but the sub-millimeter excess was clearly under-estimated (obtaining a non-reduced $\chi^2$ of $17\,000$, to be compared to $\sim 3\,500$ that we derive later on). Consequently we increased the complexity of the model by including another dust belt (that shares the same grain size distribution and $\alpha_{\mathrm{in}}$ as the main belt), and investigated two scenarios: the additional belt is either inwards or outwards of the dust belt detected with SPHERE (families ``I'' and ``O'', respectively). The hypothesis that both belts have the same grain size distribution is strong and may be unrealistic given that the dynamical timescales will be different in the two belts. Nonetheless, SED modeling is a known degenerate problem, and we opted for that solution to reduce the complexity of the modeling. For each model, we compute two SEDs, and then simultaneously find the scaling factors (from which we can determine the dust masses within the size range $s_{\mathrm{min}}$ and $5$\,mm) that best reproduce the entire SED up to mm wavelengths. The best fit models are shown in Figure\,\ref{fig:sed} (left and right, respectively), and Table\,\ref{tab:sed} summarizes the results. 

Before describing the results, a word of caution on their interpretation. One should be aware that because of the degeneracy between $s_{\mathrm{min}}$ and $r_0$, by fixing $r_0$ for one of the dust belt, we more or less constrain $s_{\mathrm{min}}$. Because we use the same $s_{\mathrm{min}}$ for both dust belts, once its value is constrained, the location of the secondary belt (inwards or outwards of the one at $25$\,au revealed by SPHERE) is also narrowed down to a small range of possible values. Because the temperature of the dust grains (for a given grain size $s$ and distance $r$) strongly depends on the absorption efficiencies, the location of the secondary dust belt that we infer also depends on the choice of optical constant.

That being said, when the ring detected with SPHERE is the inner belt (family ``O''), we find that the minimum grain size has to be much smaller compared to models of family ``I'' ($0.02$\,$\muup$m versus $3.0$\,$\muup$m, respectively), as smaller grains are usually warmer than larger grains, at a given distance to the star. Also, we can hardly constrain the location of the extra belt but it seems that it has to be much farther away than the tentative detection of the secondary ring ($\sim 300$ versus $\sim55$\,au). In fact, the best fitting model is at the boundary of the uniform prior but we did not compute additional models: the main conclusion being that the outer belt must consist of really cold dust grains. 

However, when the additional belt is inwards (family ``I''), we find that the outer slope of the ring at $25.0$\,au is extremely shallow ($\sim -0.27$ in surface density), which would suggest that it could encompass the contribution of the tentative secondary ring that we detect (and may in fact bias the determination of $\alpha_{\mathrm{out}}$). The best fit model of family ``O'' lead to a really massive disk ($4.4$\,M$_\oplus$ of dust for the outer disk, to be compared with $0.57$\,M$_\oplus$ for family ``I''). Maintaining such a large amount of small dust grains over $\sim 5-10$\,Myr is most likely challenging, as it would probably be collisionally eroded on a much shorter timescale, as more massive disks tend to evolve faster (e.g., \citealp{Wyatt2008}). Finally, we find that the $\chi^2$ is $\sim 1.5$ times smaller for the best fit model of family ``I'' compared to the best fit model of family ``O'' (for the same number of free parameters), with non-reduced $\chi^2$ values of $3\,100$ and $4\,500$, respectively. Therefore, we conclude that it is more likely that there is an extra dust belt closer in to the star, behind the coronagraph of our SPHERE observations, even though there are still some discrepant results as further discussed in Section\,\ref{sec:radpress}. This potential inner belt remains to be investigated with for instance high angular resolution observations with ALMA (Bayo et al. in prep).

\section{Discussion}
\label{sec:discussion}

Our SPHERE IRDIS DPI observations revealed a complex system around the young M2 type star TWA\,7. We resolved a radially extended main belt, at $\sim 25$\,au, a spiral arm arising from the main birth ring located on the West side, and a faint outer ring, at $52$\,au. The modeling of the SED suggests that there is an additional belt, closer in to the star at a distance of $\sim 7$\,au (its exact location depending on the dust properties used when modeling the SED).

\subsection{Comparison with the HST observations}

\citet{Choquet2016} presented HST/NICMOS near-infrared scattered light observations of TWA\,7, which were obtained the 26$^{\mathrm{th}}$ of March 1998. Their modeling results suggest that a family of models can reproduce the observations: it is unclear if the inner edge of the disk is actually resolved in the HST data, and they found that a continuous disk can also match the NICMOS data. They find a reference radius of $\sim 35-45$\,au, depending on whether the disk is continuous or a ring, an inclination between $20-30^{\circ}$, and a position angle between $50-60^{\circ}$ East of North (both values remaining loosely constrained overall).

Overall, the modeling of the SPHERE data suggests a different geometry for the disk, with a reference radius close to $25$\,au, an inclination of $13^{\circ}$, and a position angle of $91^{\circ}$. One possible explanation is that the SPHERE observations reveal the emission from the disk for all azimuthal angles, while the disk is not detected in the North-West quadrant of the HST observations. In addition, the spiral-like feature that we detect in the new observations seems to be also detected in the NICMOS observations, where it appears as a ``bar'' in the South-West quadrant. This could bias the modeling and explain the discrepancies between the different inferred geometries.

\subsection{Polarized phase function}
\label{sec:ghg}

With the availability of high-contrast, high angular resolution instruments, we are obtaining more and more constraints on the phase function of dust grains in debris disks (e.g., \citealp{Stark2014,Perrin2015,Olofsson2016,Esposito2016,Millar-Blanchaer2016,Engler2017,Milli2017}). Most of the debris disks in those studies have relatively high inclinations, as opposed to TWA\,7. With an inclination of $13^{\circ}$ for the best fit model, we are probing scattering angles in the range $90\pm13^{\circ}$. Even though we find that the value of $g$, for the Henyey-Greenstein approximation, should be close to $0.63$ (with significant uncertainties) we have no constraints on how the polarized phase function behaves at different angles, and we are not probing where it actually peaks (for $g = 0.6$ our parametric function peaks at $\theta \sim 50^{\circ}$). This is an inherent limitation that severely prevents us from better characterizing the properties of the dust grains around one of the few disks surrounding an M type star.

\subsection{Upper limits on planets from NACO and SPHERE}
\label{sec:planets}

\begin{figure*}
\centering
\includegraphics[width=0.45\hsize]{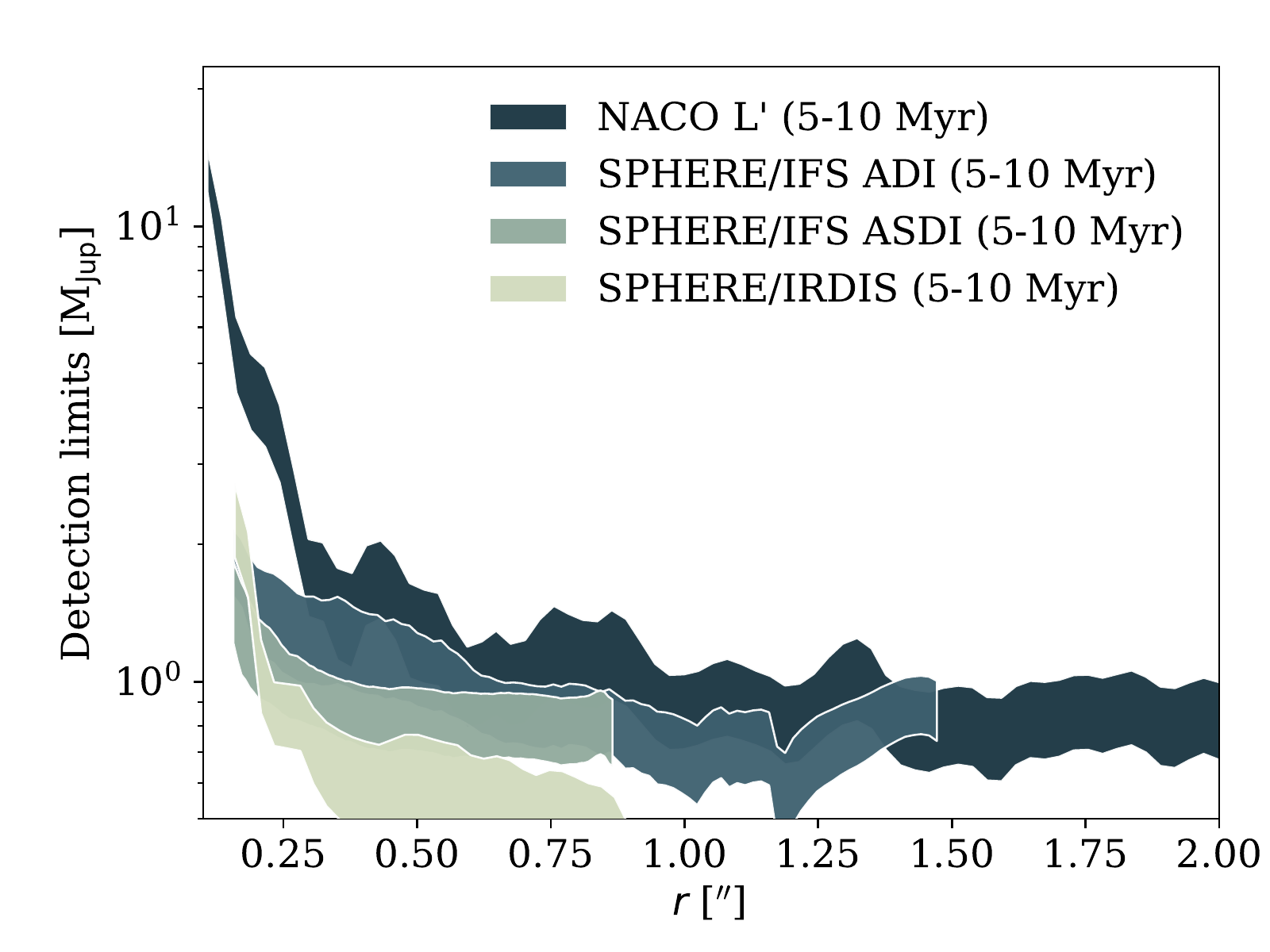}
\includegraphics[width=0.45\hsize]{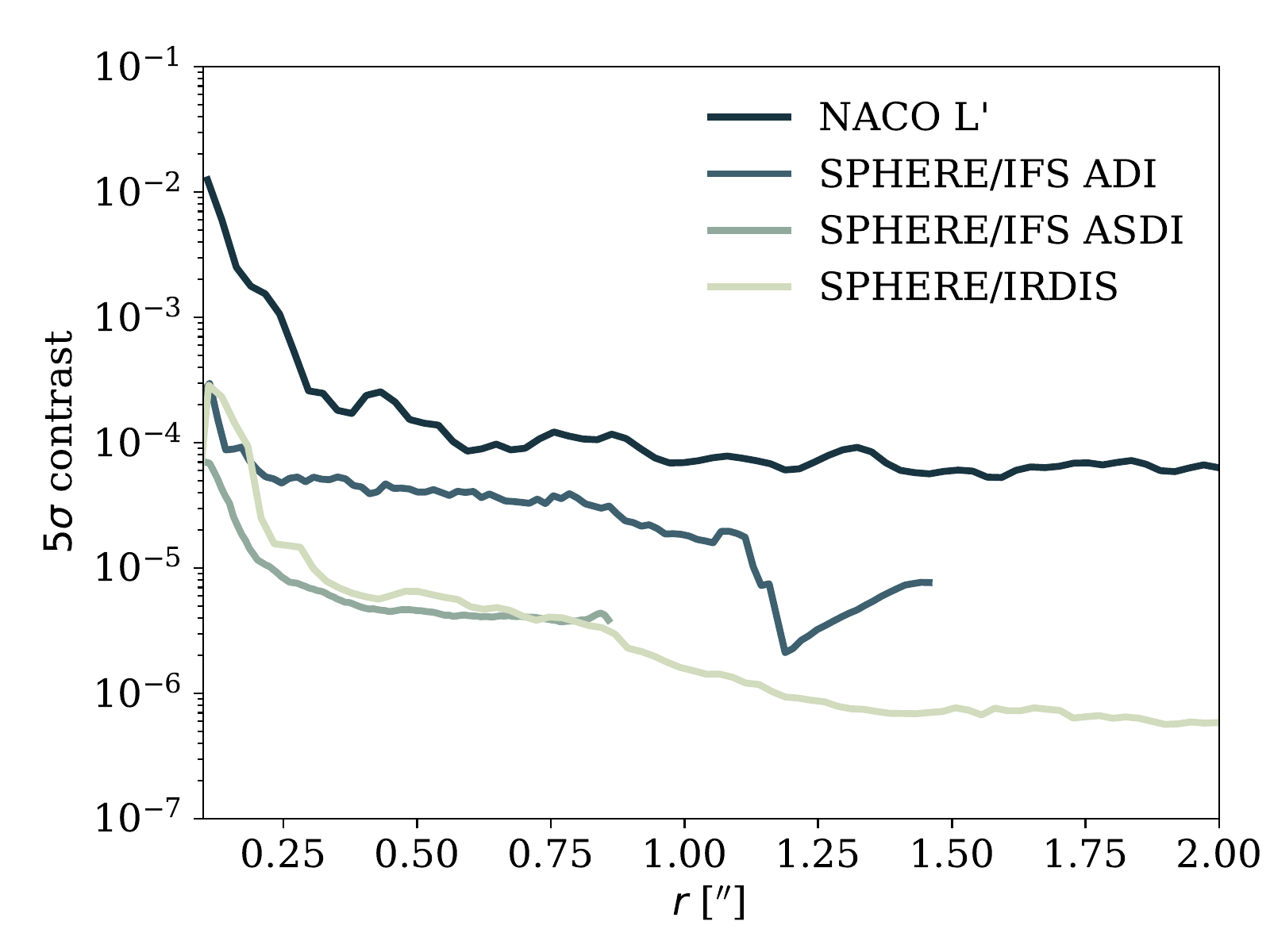}
\caption{\textit{Left panel:} upper limits on the masses of point sources in the vicinity of TWA\,7. Values below $0.5$\,M$_{\mathrm{Jup}}$ are not shown because the models are not available in this mass range. The vertical dispersion is related to the different ages assumed (between $5$ and $10$\,Myr). \textit{Right panel:} $5\sigma$ contrast detection limits.}
\label{fig:upper_limits}
\end{figure*}

Due to the youth and favorable distance of TWA\,7, the high expected brightness of young planets at near-IR wavelengths, and the good contrast performance of the AGPM and extreme adaptive optics, our NACO and SPHERE images have a deep sensitivity to planetary companions in the system. Given that no candidates are detected, beside the three objects already classified as background by \citet{Wahhaj2013,Biller2013}, we can use the achieved contrast to put a model-dependent upper limits on the masses of any wide-orbit planets around TWA\,7. We converted the absolute magnitude $5\sigma$ detection limits using COND models \citep{Baraffe2003} at ages of $5$ and $10$\,Myr. One should note that models for masses smaller than $0.5$\,M$_{\mathrm{Jup}}$ are not available, and therefore we masked out that region in Figure\,\ref{fig:upper_limits} which shows the upper limits derived from the NACO, SPHERE/IRDIS, and SPHERE/IFS observations on the left, and the $5\sigma$ contrast limits on the right panel. Although the spectral diversity of the IFS allows deeper contrast through ASDI, the detection limits are difficult to extract in that case as they rely on several strong hypotheses regarding planet spectrum (\citealp{Maire2014}, \citealp{Rameau2015}). We can confidently exclude Jovian mass planets beyond $1\arcsec$ from the central star, and between $1-2$\,M$_{\mathrm{Jup}}$ at about $0.5\arcsec$, depending on the age of the system.

\subsection{Detection of a spiral arm in a debris disk}

Having placed stringent constraints on the presence of giant planets around TWA\,7, it is interesting to discuss the possible mechanisms responsible for the spiral arm. Even though its detection remains at low S/N, the spiral-like feature is present in both SPHERE datasets, and the HST observations support the presence of an over-density in the southern region of the disk. To the best of our knowledge, this would be one of the first detection of a spiral arm in a gas-poor debris disk (HD\,141569\,A displays some interesting structures but still contains large amount of gas, \citealp{Clampin2003}, \citealp{Flaherty2016}, \citealp{Perrot2016}). Asymmetric radial structures have been detected around AU\,Mic (\citealp{Boccaletti2015}, Boccaletti et al., submitted), but the low S/N of the detection around TWA\,7, combined with the larger distance (a factor $3.5$) prevents us from distinguishing individual ``clumps'' (if there are any) from a continuous spiral over-density.

Spiral arms have been unambiguously detected in several young proto-planetary disks (see \citealp{Garufi2017} and references therein), and different processes can explain their presence, ranging from gravitational instabilities (\citealp{Lodato2004}), to shadows casted onto the outer disk (\citealp{Montesinos2016}), but these mechanisms require a large amount of gas to launch a spiral feature in the disk (see e.g., \citealp{Benisty2017}). For TWA\,7, no circumstellar gas has been detected so far (e.g., non-detection of $[\ion{O}{i}]$ with Herschel, \citealp{Riviere2013}) and based on the SED \citet{Kral2017} predicted a total gas mass lower than $10^{-4}$\,M$_\oplus$, suggesting that overall there should be little amount of second generation gas in the disk around TWA\,7. Therefore the aforementioned mechanisms are most likely unable to explain our observations. Nonetheless, gravitational interactions between an eccentric planet and the planetesimals in the debris disk can be responsible for short-lived spiral feature. \citet{Nesvold2013} showed that such a spiral can appear in their numerical simulations, but because of collisional activity the spiral arm quickly breaks up, which will later result in an apse-aligned eccentric disk. Running such detailed numerical simulations for the debris disk around TWA\,7 is out of the scope of this paper, but given the stringent upper limits we derived from both the NACO and SPHERE dataset, this scenario seems unlikely. The formation of a spiral over-density is also discussed in \citet{Pearce2015}, which we further discuss in section\,\ref{sec:spiral_gap}.

Massive collisions of planetesimals could also be responsible for the apparition of spirals in debris disks. For instance, \citet{Kral2015} noted the formation of a short-lived spiral (followed by the formation of concentric ripples) in their numerical simulations. However, the spiral's brightness dims out very rapidly according to their work, on a timescale shorter than about 100\,years, and the ripples fade away after $\sim 1000$\,years in their setup. One should note that in \citet{Kral2015}, the collision took place at about $6$\,au from the star, hence, the dynamical timescale for the dissipation of the spiral would be $\sim 16$ times longer at $20-25$\,au. Nonetheless, in our observations, the spiral appears as if it had not had the time to wrap around the star (only detected on the western side), which would suggest that the collision would have happened sometime during the past few centuries. 

An alternative scenario, which also remains speculative, is related to the spectral type of TWA\,7. As a low-mass star, it is prone to flaring events, which can redistribute the small dust grains outside of the main belt (e.g., \citealp{Augereau2006} for the case of AU\,Mic). TWA\,7 has been reported to have strong flaring events (\citealp{Uzawa2011}), and one can hypothesize that an anisotropic flare could have blown out some dust grains in a preferential direction. The potential presence of strong stellar flares and winds can have additional consequences on the disk's structure which we discuss in the next section.

\subsection{Potential impact of stellar winds on the disk}

\begin{figure*}
\centering
\includegraphics[width=\hsize]{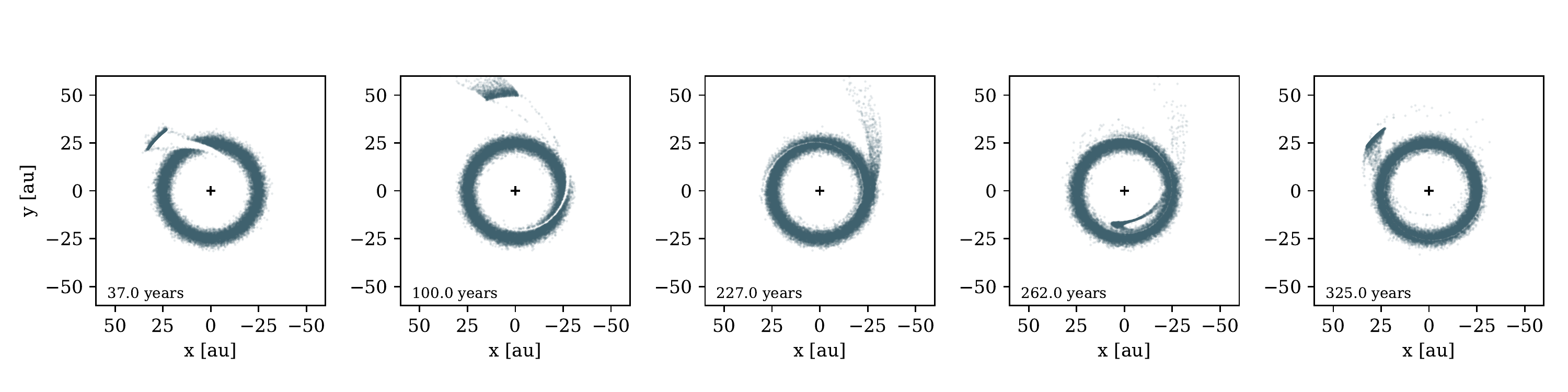}
\caption{N-body simulation of the effect of anisotropic stellar winds on an ensemble of massless particles, creating a spiral arm in the disk.}
\label{fig:stellar_winds}
\end{figure*}

Our modeling of the SPHERE/IRDIS DPI observations suggests a very flat outer slope for the volumetric density profile (in $-1.52$, resulting in a surface density profile in $-0.52$) and that there is lot of matter in the regions farther than $25$\,au. A possible explanation for this radially extended outer region is that it is made of small grains that are collisionally produced in a narrow ``birth ring'' at $25$au and placed on high-eccentricity orbits, which enter deeply in the $>25$\,au region, by radiation pressure or, more likely for a young low-mass star, stellar wind. However, \citet{Strubbe2006}, who studied the potentially stellar-wind-dominated disk around AU\,Mic, \citet{Thebault2008}, and \citet{Kral2013}, have shown that, in this case, the surface density profile beyond the ring should tend towards a $r^{-1.5}$ slope, which is much steeper than the $-0.52$ value derived in this study for TWA7. It is therefore difficult to reconcile this shallow radial profile with an underlying ring-like structure. A first possible solution would be that the S/N is relatively low, and therefore we are not constraining very well the radial profile at large separations. Additionally, because  the (tentative) secondary ring that we detect is not accounted for in our modeling, the radial profile between $0.9\arcsec$ and $1.5\arcsec$ could well be the cumulative contribution of two dust rings, and not only one, hence artificially flattening the radial profiles. Deeper observations at mid-IR wavelengths (with JWST for instance) should help to better constrain the radial profile of the disk as well as to confirm the presence of the secondary ring.

\subsubsection{A spiral arm triggered by a massive coronal mass ejection?}

Concerning the spiral arm that we detect, we can test the hypothesis that it is the consequence of a massive coronal mass ejection (CME) along the equatorial plane of the star and in one preferential direction. If the CME is collimated and intercepts a small cross-section of the disk, it could well set a fraction of particles on eccentric orbits, which would appear as a spiral arm in the disk. The CME could locally, and for a short period of time, increase the $\beta$ ratio of the small dust grains, $\beta$ being defined as
\begin{equation}
\beta = \frac{F_{\mathrm{rad}} + F_{\mathrm{wind}}}{F_{\mathrm{grav}}},
\end{equation}
as in \citet{Strubbe2006}, where $F_{\mathrm{rad}}$ is the radiation pressure force, $F_{\mathrm{wind}}$ is the force exerted by the wind on the small dust grains, and $F_{\mathrm{grav}}$ the gravitational force. For this simple exercise, we set $F_{\mathrm{rad}}$ to $0$ (radiation pressure being relatively weak for M type stars), meaning that $\beta = 0$ when there is no wind. To estimate the effect of stellar winds on the distribution of small dust grains, we performed simple N-body simulations. The central star has a mass of M$_\star = 0.5$\,M$_\odot$, and we distribute $25\,000$ massless particles with semi-major axis drawn from a normal distribution that peaks at $25$\,au (orbital period of $\sim 175$\,years), with a $2$\,au standard deviation. The initial eccentricity of each particles is set to $0$ and their locations is randomly distributed over $2\pi$ in mean anomaly. With time steps of $4$\,days, we estimate the position and velocity of the particles, using a Runge-Kutta integrator at the $4^{\mathrm{th}}$ order. We postulate that the stellar wind will act as a change in the unit-less $\beta$ ratio, and at each time step of the simulation, the particles ``feel'' a star that has a mass equal to M$_\star (1 - \beta)$.

We let the particles evolve until $t_0$, when we locally change the $\beta$ value for particles that have an azimuthal angle $\nu = \mathrm{arctan2}(y,x)$ between $[-10^{\circ}, 10^{\circ}]$. This range of values for the width of the CME was chosen so that enough particles are ejected from the main disk and that they still remain more or less collimated altogether to produce a spiral-like shape. Particles that leave this region have their $\beta$ values set back to $0$. For the time dependency, the wind exerts a force on the particles such as $\beta_\mathrm{wind} = 50$ (based on the results from \citealp{Augereau2006}), between $t_0$ and $t_0 +  \delta t$.

The point of this exercise is not to make a complete analysis of the origin of the spiral arm that we detect with our observations, but to test the viability of a scenario. We investigated the effect of each of the aforementioned parameters, and inspected the results visually. Figure\,\ref{fig:stellar_winds} shows the time evolution for a given simulation, over a span of several hundreds years. For this example, we took $t_0 = 2$\,years and $\delta t = 0.3$\,years. Shortly after $t = t_0$ a spiral, devoid of particles, appears in the disk (where the particles originally were), but its width decreases for each orbit owing to Keplerian shear. On the other hand, the grains that experienced the wind are set on an elliptical orbit once the wind has stopped (and provided that $\beta_{\mathrm{wind}}$ is not too strong), with their apocentre farther than the original radius of the disk. All these particles tend to travel all-together, therefore, a spiral arm appears in the image at two distinct moments per orbit: when the group of particles is entering or leaving the birth ring (middle and rightmost panels of Fig.\,\ref{fig:stellar_winds}). In our simulations, given that collisions are not taken into account, this phenomenon will persist over time. Nonetheless, if the optical depth is low enough, the train of particles may not collide too often with other particles of the birth ring, and thus the spiral could be long-lived. A movie of the simulation can be found at \url{https://goo.gl/MqiFCh}.

Nonetheless, the $\delta t$ timescale is far too long compared to the stellar rotation period for the CME to stay in one preferential direction ($0.3$\,years compared to $\sim5$\,days, \citealp{Watson2006}). The stellar wind would most likely sweep across the entire disk multiple times. If we require $\delta t$ to be much lower than a typical stellar rotation period, or of the order of the typical duration of the flaring event reported by \citet{Uzawa2011}, then the corresponding $\beta_{\mathrm{wind}}$ value becomes much higher. As an example, for $\delta t = 0.5$\,days, we find that a similar spiral arm requires $\beta_{\mathrm{wind}}=7500$, a value that seems unrealistically high. Indeed, $\beta_{\mathrm{wind}}$ is directly proportional to stellar mass loss rate and inversely proportional to grain sizes following the relation (\citealp{Augereau2006}):
\begin{equation}\label{eqn:bwind}
\beta_{\mathrm{wind}} = \frac{3}{32\pi}\frac{\dot{M}_\star v_\mathrm{sw}C_\mathrm{D}}{G M_\star \rho s},
\end{equation}
with $C_\mathrm{D}$ a factor close to $2$, $\dot{M}_\star$ the stellar mass loss rate, $v_\mathrm{sw}$ the speed of the stellar wind, $G$ the graviational constant, and $\rho$ the dust density. Even though TWA\,7 is younger than AU\,Mic, and therefore potentially more active, a $\beta_{\mathrm{wind}}=7500$ value seems rather improbable, as it would require extremely small dust grains (which would not scatter the stellar light efficiently in $H$-band) or an extremely large mass loss rate. For AU\,Mic, \citet{Augereau2006} found $\dot{M}_\star \sim 47 \dot{M}_\odot$ in quiescent state,  $\sim 2450 \dot{M}_\odot$ during flares (leading to $\beta_{\mathrm{wind}} \sim 40-50$), which averaged to  $\sim 300 \dot{M}_\odot$ (assuming flares occur $10$\% of the time). This means that the flares of TWA\,7 should be $7500/50 = 150$ times stronger according to our results. Furthermore, it would have to remain a unique event, as if there had been similar CMEs in the past centuries, we would not observe only one spiral arm, but several others. Overall, while this scenario is interesting, as it is related to specific properties of low mass stars, it remains highly improbable that the spiral arm is the sole consequence of a massive CME from the central star.

\subsubsection{The blow-out size due to stellar winds}
\label{sec:radpress}

\begin{figure}
\centering
\includegraphics[width=\hsize]{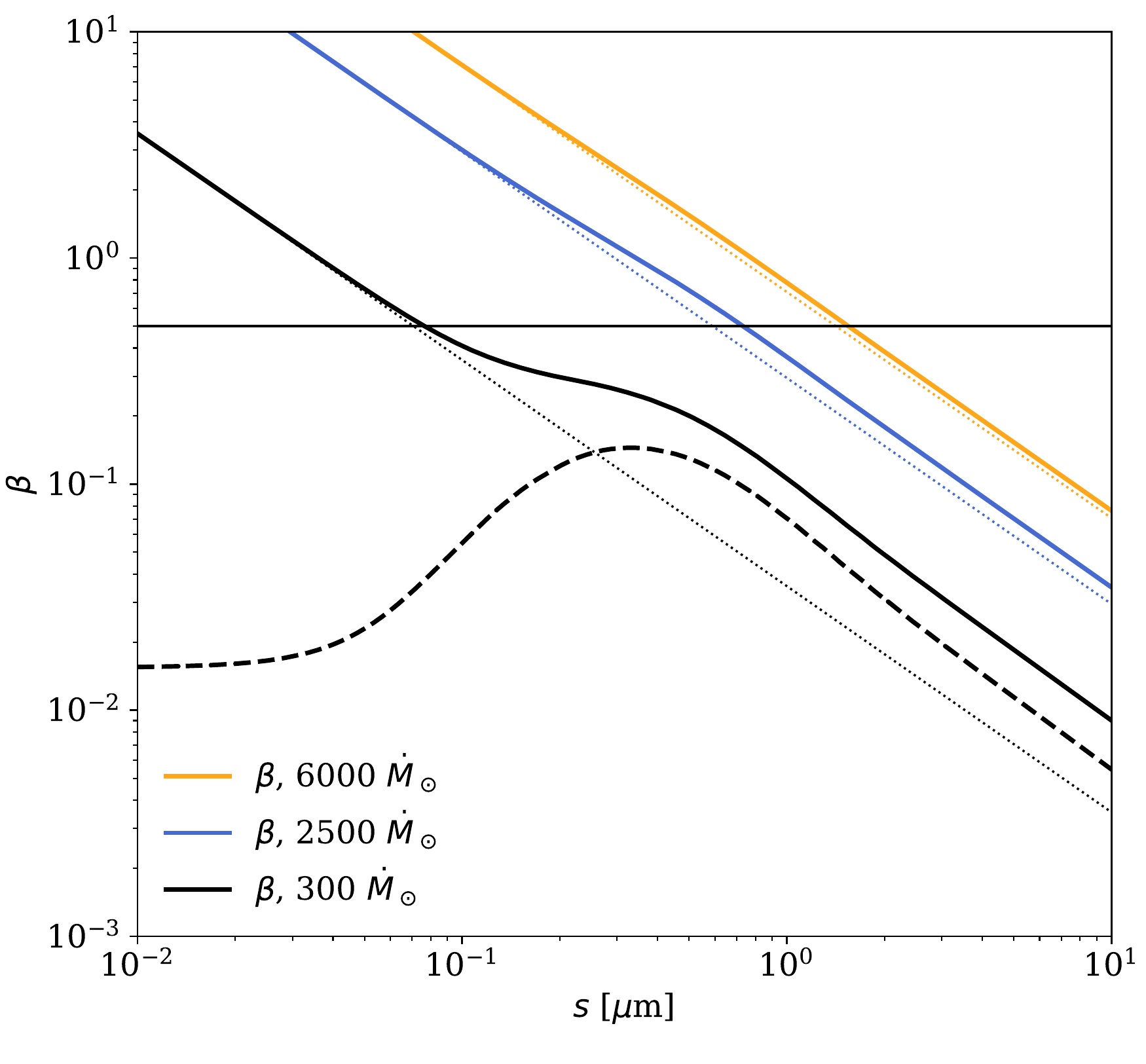}
\caption{$\beta$ ratio (solid lines), including the effect of stellar winds (dotted lines) and radiation pressure (dashed line), as a function of the grain size, for several values of the stellar mass-loss rate.}
\label{fig:beta_wind}
\end{figure}

In Section\,\ref{sec:sed} we found that the best-fitting model has a minimum grain size of about $3$\,$\muup$m, which is a fairly large value for a low-mass star such as TWA\,7 given that the effect of radiation pressure is expected to be negligible. Figure\,\ref{fig:beta_wind} shows the $\beta$ ratio as a function of the grain size for the astro-silicates composition used when modeling the SED. On top of the radiation pressure (dashed line) we included the contribution of stellar winds (dotted lines) for several mass-loss rates (following Eqn.\,\ref{eqn:bwind}). To reach a blow-out size of a few microns (the size for which $\beta = 0.5$), it appears clearly that the average mass-loss rate has to be greater than $\sim 5\,000\,\dot{M}_\odot$. As mentioned before, \citet{Augereau2006} reported an average mass-loss rate of about $300\,\dot{M}_\odot$ for AU\,Mic. It is therefore difficult to reconcile the minimum grain size of $3$\,$\muup$m inferred from the SED modeling with the combined effects of radiation pressure and stellar winds. A possible work-around would be to have less ``dirty'' grains, that would be more transparent in the optical and near-infrared. At the same distance from the star, such grains would be slightly colder than the astro-silicate grains we used, which would in turn decrease the minimum grain size required to reproduce the infrared excess.

\subsection{Inner and outer dust belts from SED modeling}

In this sub-section, we refer to the inner and outer belts inferred from modeling the SED, i.e., the belts at $\sim 7$ and $25$\,au. The fact that the debris disk around TWA\,7 is best modeled by multiple dust belts may be an indicator of the presence of planets in between the belts. Indeed, a planet can efficiently remove dust grains along its orbit. The size of the dust-depleted region, the so-called chaotic zone, follows
\begin{equation}
\Delta a \propto \mu^{\alpha} a_p e_p^{\beta},
\end{equation}
(\citealp{Lazzoni2017}, and references therein), where $\mu$ is the mass ratio between the planet and the host star, $a_\mathrm{p}$ and $e_\mathrm{p}$ are the semi-major axis and eccentricity of the planet, respectively, and $\alpha$ and $\beta$  can vary depending on the formalism adopted.

Following the procedure described in \cite{Lazzoni2017}, we calculate the mass of a putative planet on circular obit, coplanar with the disk, responsible for the gap between the innermost belt, placed at $6.8$\,au, and the second belt, placed at $25$\,au (Family ``I'' in section\,\ref{sec:sed}). Assuming a stellar mass of $0.55$ $M_{\odot}$, we find that the inferred geometry of the disk can be explained by a companion of $31$\,M$_\mathrm{Jup}$ that orbits at $\sim 15$\,au ($\sim 0.43\arcsec$, using Eqns\,3 and 4 of \citealp{Lazzoni2017}). However, comparing these values with the detection limits of TWA\,7, such a massive companion would have easily been detected. When considering eccentric orbits, the situation slightly improves, with less massive companions for increasing values of $e_\mathrm{p}$. However, in order to have a planet smaller than few Jupiter masses, we have to consider values of $e_\mathrm{p} \sim 0.3$ and such eccentric orbits should induce an eccentricity on the disk itself, on timescales of a few $1\,000$\,yrs \citep{Mustill2009}. For this reason a single planet seems to be unlikely to be responsible for sculpting the gap between the two dust belts.

We can then consider more than one planet in the region devoid of dust grains. In that case, we also have to introduce a dynamical stability criterion: if the planets get too close to each other one of them may be scattered away. In the following, we consider a compact system, at the dynamical stability limit, with two and three equal-mass planets on circular orbits.

For the two planets case, solving Eqn\,29 of \citet{Lazzoni2017}, we find a mass of M$_\mathrm{p}=3.5$\,M$_\mathrm{Jup}$ and semi-major axis $a_1=9.8$\,au and $a_2=18.5$\,au for the inner and outer planets, respectively. For the three planets case, in line with the findings of \citet[][with a cumulative occurence rate of $\sim 2.5$ planets per M dwarf]{Dressing2015}, we obtain a mass of M$_\mathrm{p}=0.2$\,M$_\mathrm{Jup}$ and semi-major axis $a_1=8.1$\,au, $a_2=13.4$\,au and $a_3=22.1$\,au for the inner, mid and outer planets, respectively. This mass is the mean value when propagating the uncertainties on the location of the belts (set to $\pm\,1$\,au). While the two planets case is ruled out by our observations (Fig.\,\ref{fig:upper_limits}), the three planets case is definitively under detection limits. One could also consider companions of different masses in order to have a bigger planet in the innermost regions and one or two smaller ones placed further out. Overall, if planets are responsible for opening a cavity in the debris disk, we conclude that there should be several low-mass (most likely sub-Jovian) planets that remain undetectable with current observations.

We note that the procedure by \cite{Lazzoni2017} does not account for the possible effect of radiation pressure or stellar wind, which, coupled to collisional activity in the disk, could attenuate the gap-inducing power of one or several putative planets for the micron-sized grains that SPHERE is preferentially probing (\citealp{Thebault2012}). This potential effect would further increase the planetary masses required to produce a given gap, thus making the single-planet scenario even more unlikely. Likewise, in the 3-planet scenario, the planet masses should probably be considered as lower limits

\subsection{Inner and outer dust belts as seen with SPHERE}\label{sec:spiral_gap}

In this sub-section, we refer to the two belts that we detect in the SPHERE IRDIS DPI image, i.e., the belts at $25$ and $52$\,au. \citet{Pearce2015} discussed the effect of an eccentric planet on a debris disk, and concluded that under certain circumstances, the original continuous debris disk could display two different rings, very similar to what we see around TWA\,7. If the mass of the planet is comparable to the mass of the disk (or larger), the evolution of the disk can follow a different path. \citet{Pearce2015} highlight different main phases of the disk's evolution, which includes the formation of a spiral wave due to secular interactions and a slow damping of the planet's eccentricity due to the planet scattering planetesimals that are crossing its orbit. As time goes by, the dust surface density starts displaying two distinct peaks. The one closer to the central star encompasses the initial pericentre and apocentre of the planet (hence the planet is not located in the gap as naively expected but rather in the inner disk) and a second peak that moves farther out, corresponding to the spiral arm. In their numerical simulations, depending on the timescale for the planet circularization compared to the secular one, the spiral may sustain over time.

\citet{Pearce2015} applied their numerical simulations to the debris disk around HD\,107146, aiming at reproducing the ALMA observations presented in \citet{Ricci2015}. They can best match the ALMA data in $19$\,Myr with a $100$\,M$_\oplus$ planet with a semi-major axis of $a = 40$\,au, an eccentricity of $e = 0.4$ that interacts with a disk which inner radius is set at $50$\,au. Even though HD\,107146 is about twice more massive (spectral type G2V), the main disk around TWA\,7 is twice closer. Therefore it is not unlikely that a (few) $100$\,M$_\oplus$ planet hides within the main dust belt, remaining undetected with SPHERE, and being responsible for the spiral arm and secondary ring at 52\,au. One should note the striking similarity between the surface density profiles of Figure\,\ref{fig:radial} and the one of Fig.\,4 of \citet{Pearce2015}. Concerning the mass of the debris disk, the results from the SED modeling (Table\,\ref{tab:sed}) only account for grain sizes up to $5$\,mm. Extrapolating the integrated grain size distribution to $100$ km-sized boulders leads to a total disk mass of $\sim82.6$\,M$_\oplus$.

The main challenge of this scenario remains the formation of an initially eccentric ($e \sim 0.4-0.6$) planet with a mass of $\sim 100$\,M$_\oplus$, at $\sim 25$\,au. A possible explanation would be that such a planet did not form in-situ, but would have formed closer in and has been scattered by a third body of at least comparable mass. 

\subsection{Tentative detection of a dusty cloud at 3.9 au}
\label{sec:coincidence}

\begin{figure}
\centering
\includegraphics[width=\hsize]{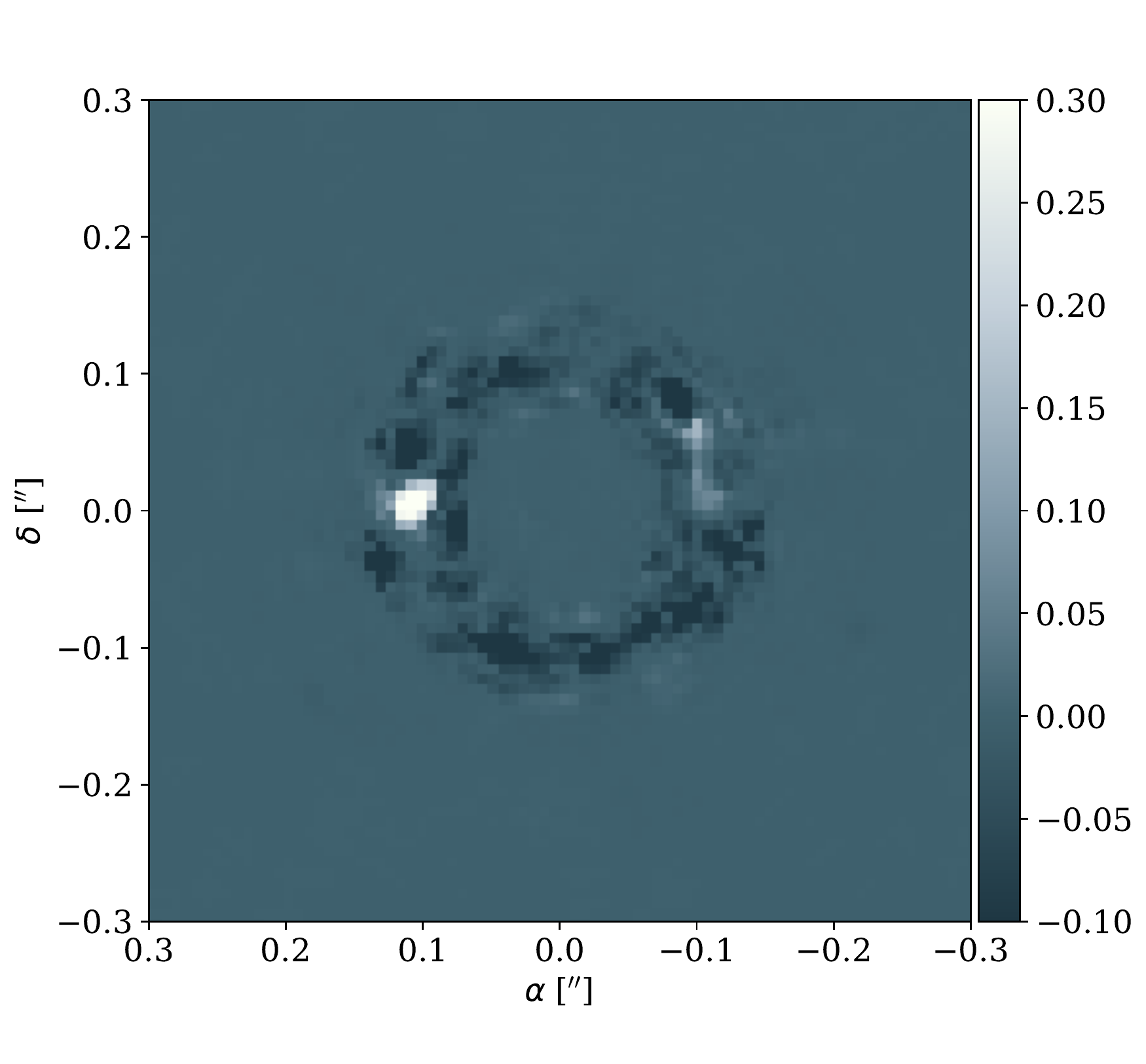}
\caption{Coincidence map from the two SPHERE/IFS epochs.}
\label{fig:coincidence}
\end{figure}

Observations at different epochs can help unveiling features that would remain undetected otherwise. We used the two epochs of SPHERE/IFS observations (2015 and 2017) and the ASDI reduction was performed using PCA, with $10$, $25$, $50$,  $100$, and $150$ components. Within concentric rings with widths of $0.1\arcsec$ we kept the reduction yielding the best contrast (with $50$ modes for both epochs), taking attenuation into account. Overall, the residuals average at zero. The image was then divided in $13$ concentric annuli and for the 2017 epoch each annulus was rotated to compensate for Keplerian motion at middle separation, with respect to the 2015 epoch. This is made possible given the almost face-on geometry of the disk. We then computed a coincidence map, by multiplying the two images pixel-by-pixel (choosing that when both pixels are negative, the resulting value is the inverse of the product). Thanks to this ``Time Differential Imaging'' the noise cancels out, while astrophysical signal is boosted. The coincidence map is shown in Figure\,\ref{fig:coincidence}.

A point source clearly appears East to the star, at a separation of $0.115\arcsec$, ($3.9$\,au), close to the edge of the coronagraph. For the Keplerian motion, we assumed a distance of $34.5$\,pc, clockwise motion (the point source disappears if the motion is counter-clockwise), and different stellar masses. We find that the results are robust within the range M$_\star = 0.16$ and $1.34$\,M$_\odot$, and the structure appears strongest for a stellar mass of $0.71$\,M$_\odot$. On the individual epoch, we find separation of $0.115\arcsec$ for both epochs, position angles of $89.6^{\circ}$ and $66.3^{\circ}$ at Julian dates $57151.97$ $57791.25$ (for 2015 and 2017, respectively).

From the 2017 epoch, we extracted a contrast spectrum  of the point source (i.e., divided by the stellar spectrum), which appears relatively flat with a median value of $7.65 \times 10^{-5}$. For the ASDI reduction, this contrast is compatible with a $\sim 5\sigma$ detection ($6.5 \times 10^{-5}$ measured at $0.115\arcsec$ from the right panel of Fig.\,\ref{fig:upper_limits}).

There are several possible explanations for the presence of this point source in the coincidence map. First, we are seeing two speckles that unfortunately aligned when de-rotating to compensate for Keplerian motion. This can be tested against by re-observing TWA\,7 and compare the results when combining all the different datasets. If the signal is arising from an astrophysical object, it is extremely unlikely that it comes from a stellar photosphere as the contrast spectrum being flat. The contrast as a function of wavelength suggests that we are most likely seeing stellar light scattered by a dusty cloud. This dusty cloud could either be self-gravitating (e.g., filling the Roche lobe of a planet, in the earliest phases) or not (e.g., an evaporating comet or a collision between planetesimals). At this point in time, it is premature to further discuss the possible origin of this intriguing feature, as it should be confirmed by new observations first. Nonetheless, this echoes the structures observed around AU\,Mic (\citealp{Boccaletti2015}) even though the one around TWA\,7 follows Keplerian motion (by construction).

\section{Conclusion}

With the advent of high contrast instruments such as VLT/SPHERE, we are making significant progress on the characterization of otherwise elusive debris disks around low mass stars. Such observations are challenging our understanding of debris disks evolution (e.g., \citealp{Boccaletti2015,Chiang2017,Sezestre2017} for the intriguing case of AU\,Mic). In this paper, we presented new VLT/SPHERE observations of the debris disk around TWA\,7, which illustrate how powerful polarized intensity observations can be, to even detect such faint debris disks seen almost face-on.

We found that the radial profile peaks at about $25.1$\,au ($0.72\arcsec$), and is very shallow at larger distances. Modeling the SED of the debris disk, we concluded that an additional belt at $\sim 7$\,au ($0.2\arcsec$) from the star can best reproduce the observed thermal emission. We report the tentative detections of a secondary ring at about $52$\,au ($1.5\arcsec$), a spiral arm that seems to originate from the main ring, and a clumpy structure at $0.11\arcsec$ ($\sim 3.9$\,au) that could be a concentration of small dust grains. We did not detect any giant planets which could have (partially) helped explain the spiral arm, reaching (sub-)Jovian mass upper limits at separations $\sim1\arcsec$ and farther out. We investigated several scenarios to explain the observed features, and the most straightforward scenario (requiring least free parameters) is that an (originally) eccentric planet with a mass comparable to the one of the disk has been sculpting it via secular interactions. If the yet-undetermined mechanism responsible for launching the ``ripple-like'' features around AU\,Mic is the same that is responsible for the spiral arm of TWA\,7 (which is also only observed on one side of the disk), then the face-on geometry of TWA\,7 is more favorable to better characterize it.

\begin{acknowledgements}
We would like to thank the anonymous referee for the valuable feedback we received, especially regarding the uncertainties derived for the best-fitting model. This research has made use of the SIMBAD database (operated at CDS, Strasbourg, France) and makes use of VOSA, developed under the Spanish Virtual Observatory project supported from the Spanish MICINN through grant AyA2011-24052. This research made use of Astropy, a community-developed core Python package for Astronomy (\citealp{Astropy}), as well as the TOPCAT software (\citealp{Taylor2005}).
SPHERE is an instrument designed and built by a consortium consisting of IPAG (Grenoble, France), MPIA (Heidelberg, Germany), LAM (Marseille, France), LESIA (Paris, France), Laboratoire Lagrange (Nice, France), INAF–Osservatorio di Padova (Italy), Observatoire de Gen\`eve (Switzerland), ETH Zurich (Switzerland), NOVA (Netherlands), ONERA (France) and ASTRON (Netherlands) in collaboration with ESO. SPHERE was funded by ESO, with additional contributions from CNRS (France), MPIA (Germany), INAF (Italy), FINES (Switzerland) and NOVA (Netherlands).  SPHERE also received funding from the European Commission Sixth and Seventh Framework Programmes as part of the Optical Infrared Coordination Network for Astronomy (OPTICON) under grant number RII3-Ct-2004-001566 for FP6 (2004–2008), grant number 226604 for FP7 (2009–2012) and grant number 312430 for FP7 (2013–2016). We also acknowledge financial support from the Programme National de Plan\'etologie (PNP) and the Programme National de Physique Stellaire (PNPS) of CNRS-INSU in France. This work has also been supported by a grant from the French Labex OSUG@2020 (Investissements d'avenir – ANR10 LABX56). The project is supported by CNRS, by the Agence Nationale de la Recherche (ANR-14-CE33-0018). It has also been carried out within the frame of the National Centre for Competence in Research PlanetS supported by the Swiss National Science Foundation (SNSF). MRM, HMS, and SD are pleased to acknowledge this financial support of the SNSF. Finally, this work has made use of the the SPHERE Data Centre, jointly operated by OSUG/IPAG (Grenoble), PYTHEAS/LAM/CESAM (Marseille), OCA/Lagrange (Nice) and Observatoire de Paris/LESIA (Paris). We thank P. Delorme and E. Lagadec (SPHERE Data Centre) for their efficient help during the data reduction process. 
J.~O., A.~B., M.~R.~S., C.~C., and N.~G. acknowledge support from the ICM (Iniciativa Cient\'ifica Milenio) via the Nucleo Milenio de Formación planetaria grant. J.~O acknowledges support from the Universidad de Valpara\'iso and from Fondecyt (grant 1180395). C.~C. acknowledges support from project CONICYT PAI/Concurso Nacional Insercion en la Academia, convocatoria 2015, folio 79150049. R.~A.-T. gratefully acknowledges funding from the Knut and Alice Wallenberg foundation. Q.~K. acknowledges funding from STFC via the Institute of Astronomy, Cambridge Consolidated Grant. MRS thanks for support from Fondecyt (grant 1141269). R.~G., C.~L., S.~D., and D.~M. acknowledge support from the "Progetti Premiali" funding scheme of the Italian  Ministry  of  Education,  University,  and  Research. This work has been supported by the project PRIN-INAF 2016 The Cradle of Life - GENESIS- SKA (General Conditions in Early Planetary Systems for the rise of life with SKA). N.~G. acknowledges grant support from project CONICYT-PFCHA/Doctorado Nacional/2017 folio 21170650. F.~M, C.~P., and M.~L. acknowledge funding from ANR of France under contract number ANR-16-CE31-0013. C.~P. acknowledges funding from the Australian Research Council (ARC) under the Future Fellowship number FT17010004. 

\end{acknowledgements}

\bibliographystyle{aa}

\appendix
\section{Observations}

\begin{figure}
\centering
\includegraphics[width=\hsize]{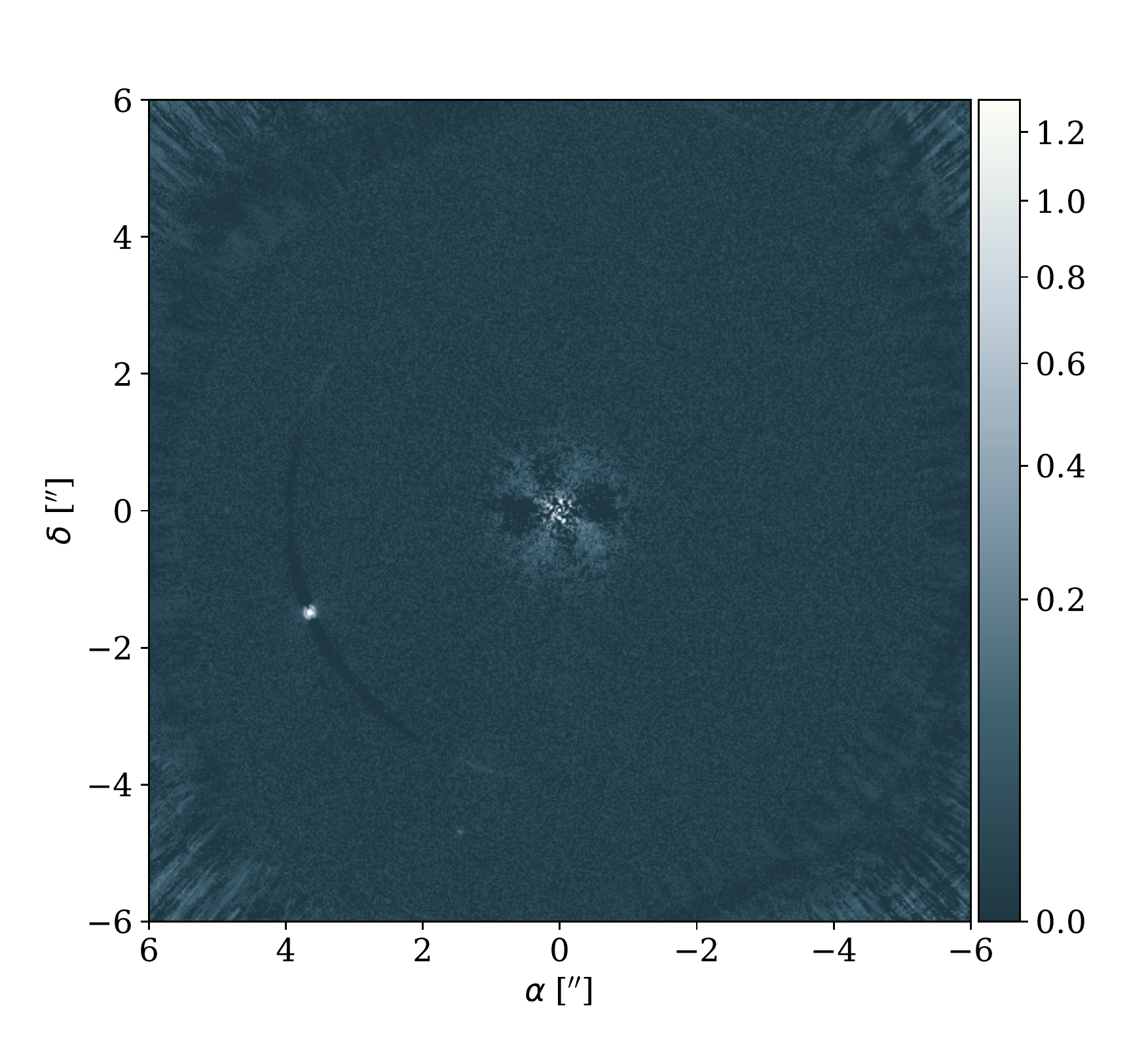}
\caption{SPHERE/IRDIS $H2H3$ image, with a square root scaling.}
\label{fig:irdis_image}
\end{figure}

\begin{figure}
\centering
\includegraphics[width=\hsize]{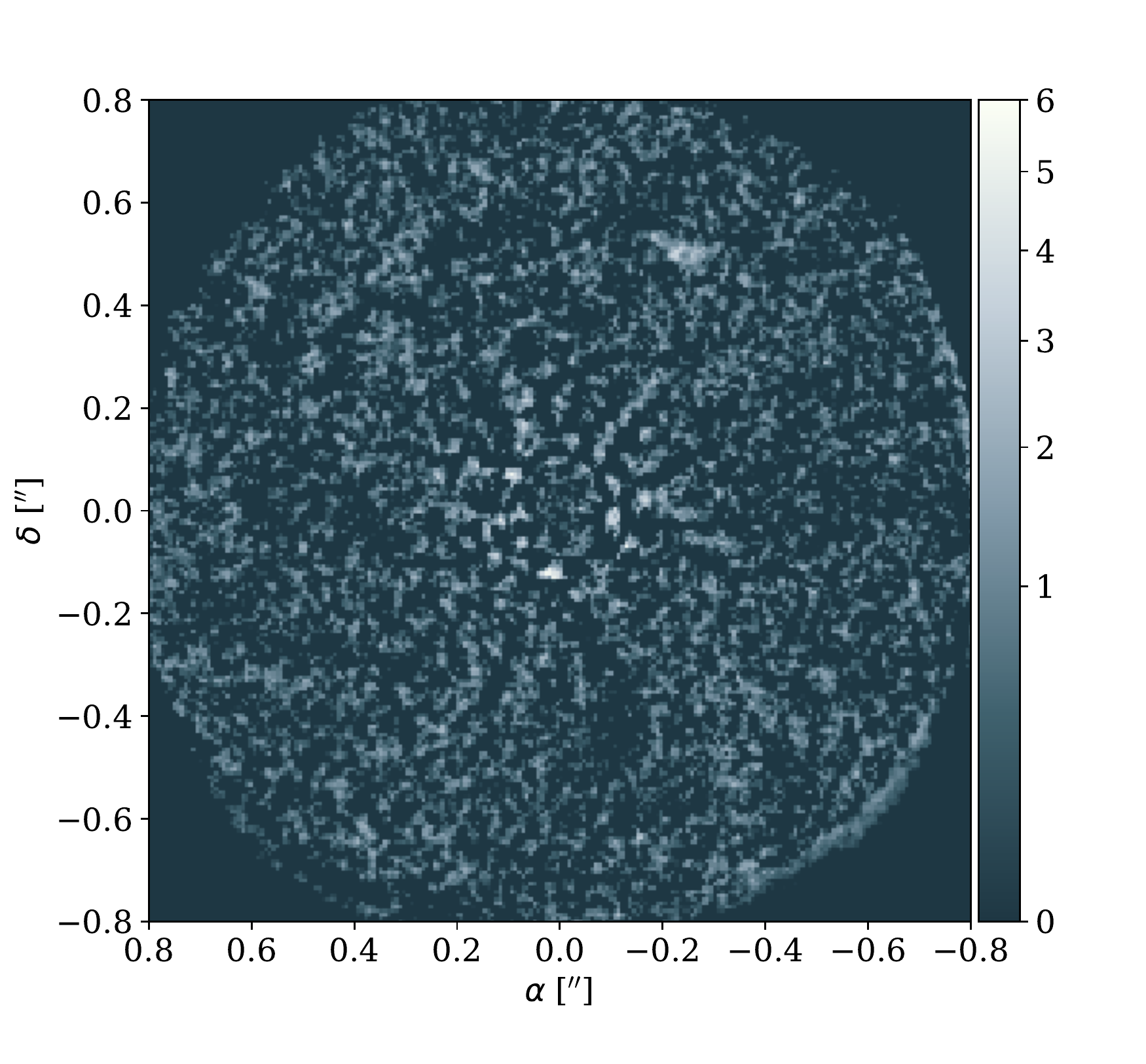}
\caption{SPHERE/IFS $YJ$ image, with a square root scaling.}
\label{fig:ifs_image}
\end{figure}

\begin{figure}
\centering
\includegraphics[width=\hsize]{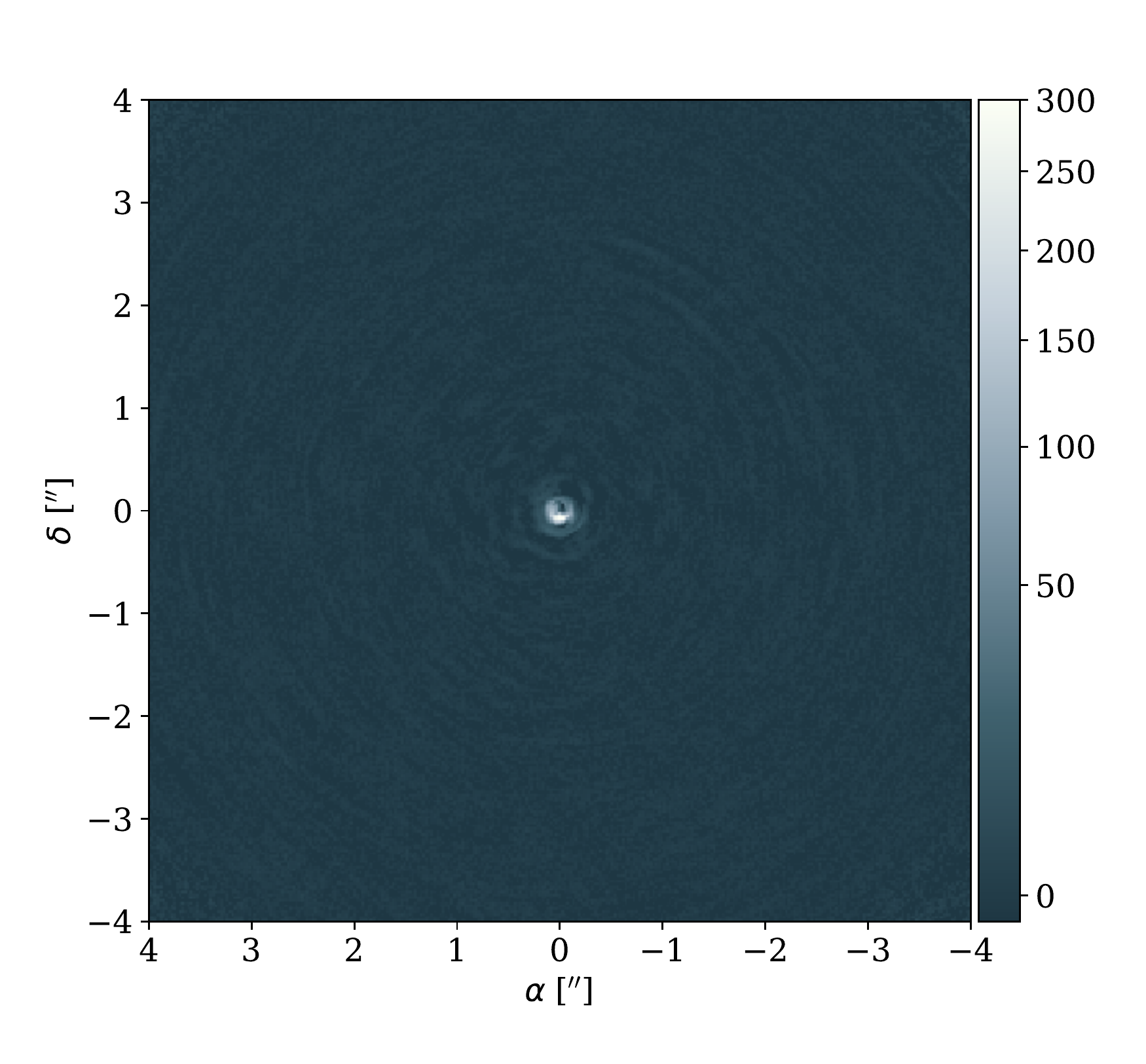}
\caption{NACO $L'$ image, with a square root scaling.}
\label{fig:naco_image}
\end{figure}

\section{Miscealleneous}

\begin{figure*}
\centering
\includegraphics[width=\hsize]{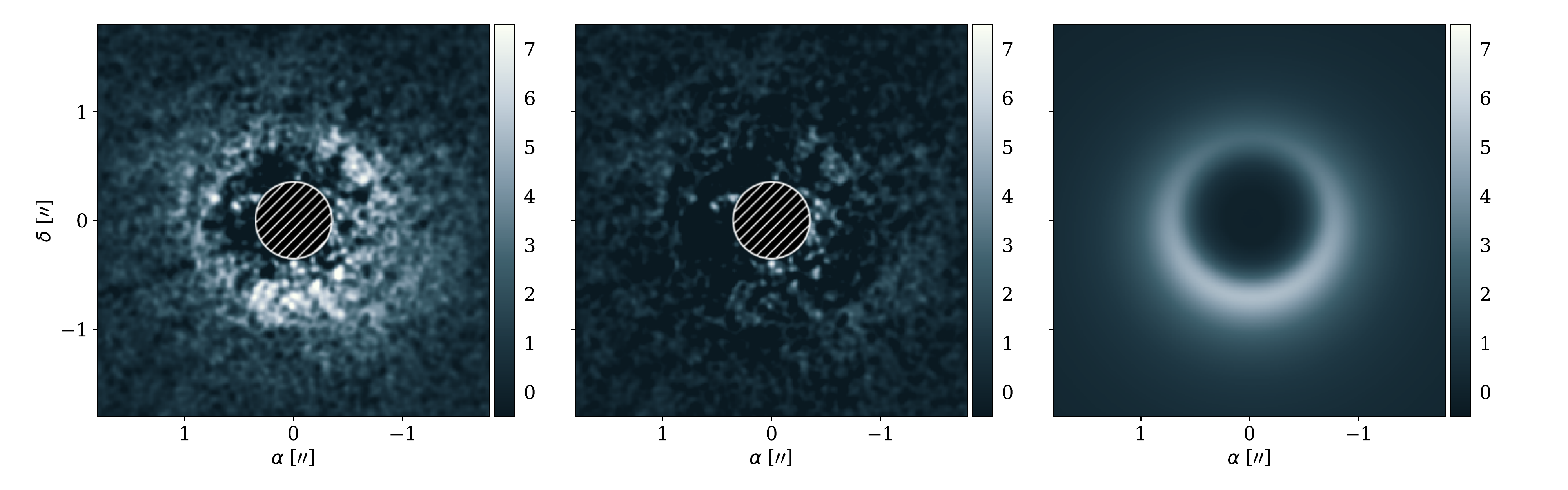}
\caption{Same as Fig.\,\ref{fig:results} but for the P97 $J$-band dataset, with the same input parameters.}
\label{fig:results_p97}
\end{figure*}

\end{document}